\documentclass[12pt,preprint]{aastex}
\usepackage[numberedappendix]{emulateapj5}
\shorttitle{NUCLEAR MORPHOLOGY IN THE TOOMRE SEQUENCE}
\shortauthors{LAINE ET AL.}
\begin{document}

\submitted{TO BE PUBLISHED IN THE DECEMBER 2003 ISSUE OF THE ASTRONOMICAL JOURNAL}
\title{{\it HUBBLE SPACE TELESCOPE}/WFPC2 INVESTIGATION OF THE NUCLEAR 
MORPHOLOGY IN THE TOOMRE SEQUENCE OF MERGING GALAXIES\altaffilmark{1}}

\author{Seppo Laine}
\affil{{\it SIRTF} Science Center, Mail Code 220-6, California Institute of 
Technology, Pasadena, CA 91125}
\email{seppo@ipac.caltech.edu}

\author{Roeland P. van der Marel, J\"{o}rn Rossa and Torsten
B\"{o}ker\altaffilmark{2}}
\affil{Space Telescope Science Institute, 3700 San Martin Drive, Baltimore, MD
21218}
\email{marel@stsci.edu, jrossa@stsci.edu, boeker@stsci.edu}

\author{J. Christopher Mihos\altaffilmark{3}}
\affil{Department of Astronomy, Case Western Reserve University, 10900 Euclid
Avenue, Cleveland, OH 44106}
\email{hos@burro.astr.cwru.edu}

\author{John E. Hibbard}
\affil{National Radio Astronomy Observatory, 520 Edgemont Road, 
Charlottesville,VA 22903-2475}
\email{jhibbard@nrao.edu}

\and

\author{Ann I. Zabludoff}
\affil{Steward Observatory, University of Arizona, 933 North Cherry Avenue, 
Tucson, AZ 85721-0065}
\email{azabludoff@as.arizona.edu} 

\altaffiltext{1}{Based on observations made with the NASA/ESA Hubble
Space Telescope, obtained at the Space Telescope Science Institute, which is 
operated by the Association of Universities for Research in Astronomy, Inc., 
under NASA contract NAS 5-26555. These observations are associated with 
proposal \#8669.}
\altaffiltext{2}{On assignment from the Space Telescope Division of the 
European Space Agency}
\altaffiltext{3}{NSF Career Fellow and Research Corporation Cottrell Scholar.}

\begin{abstract}

We report on the properties of nuclear regions in the Toomre Sequence of
merging galaxies, based on imaging data gathered with the {\it Hubble Space
Telescope} WFPC2 camera. We have imaged the 11 systems in the proposed 
evolutionary merger sequence in the F555W and F814W  broad-band filters, and 
in H$\alpha$+[\ion{N}{2}] narrow-band filters. The broad-band morphology of 
the nuclear regions varies from non-nucleated starburst clumps through 
dust-covered nuclei to a nucleated morphology. There is no unambiguous trend 
in the morphology with merger stage. The emission-line morphology is extended 
beyond the nucleus in most cases, but centrally concentrated (within 1 kpc)
emission-line gas can be seen in the four latest-stage merger systems. We have
quantified the intrinsic luminosity densities and colors within the inner
100~pc and 1~kpc of each identified nucleus.  We find little evidence for a
clear trend in nuclear properties along the merger sequence, other than a
suggestive rise in the nuclear luminosity density in the most evolved members
of the sequence. The lack of clear trends in nuclear properties is likely due
both to the effects of obscuration and geometry, as well as the physical
variety of galaxies included in the Toomre Sequence.

\end{abstract}


\section{INTRODUCTION}
\label{s:intro}

Disk galaxy mergers are believed to be responsible for triggering a variety of
global {\it and} nuclear responses in galaxies. The global effects have been
well documented, both with observations, starting with Zwicky's extensive work
(1950, 1956, 1964), and with numerical simulations, first convincingly 
demonstrated by \citet{tmr72}. Work during the last decades  strongly suggests
that  a fraction of elliptical galaxies has formed as a result of disk galaxy
merging \citep*[see][and references therein]{sch98}. The morphology of tidal
tails has been used to trace the mass distribution in galactic halos
\citep*{dub96,hos98,spr99}, and the tails themselves are possible birthplaces
of  some dwarf galaxies \citep{bar92,elm93,duc01}.  Mergers have also been
closely connected to luminous and ultraluminous infrared galaxies
\citep*{lons84,jos85,schw90,san88a,sco00,bor00}. 

Perhaps the most dramatic physical process associated with disk galaxy merging
is the inflow of gas into the nuclear region and the consequent excitation of
nuclear starburst and AGN activity. This process has been reported from
simulations \citep{bar91,hos94,bar96} and from observations
\citep*{jos85,san88a,hn92,bah95,sur01}. Whether the gas ``hangs up'' and forms
stars in the inner kiloparsec, or continues to flow inward towards a putative
AGN, has a strong impact on the luminosity and evolution of the merger. The
details of the nuclear gasdynamics will depend on the structure of the host
galaxies and the dynamical stage of the interaction \citep{hos96,bar96}. 
Several scenarios have been suggested in which interactions evolve from
starburst-dominated to AGN-dominated regimes as the galaxies merge
\citep{weed83,san88a,vlx95}.

Unfortunately, until now, both observations and numerical simulations have
lacked the spatial resolution needed to study the evolution of the merging
nuclei on scales smaller than a few hundred parsecs. For this reason, our
understanding of how mergers fuel nuclear starburst and AGN activity, and drive
galaxy evolution from the nucleus out, has remained woefully incomplete.
Theoretical arguments suggest that the evolution of the merging nuclei is where
the merger hypothesis for the formation of elliptical galaxies through disk
galaxy mergers faces its most stringent test. 

A wide variety of physical processes may shape the nuclei. Purely stellar
dynamical merging alone would tend to produce diffuse nuclei with large cores
and a very shallow nuclear surface brightness gradient \citep{her92} unless the
progenitor nuclei were dense \citep{bar88}. In the case of gas-rich disk
galaxies, the dissipative flow of gas into the nuclei and accompanying star
formation would tend to result in a steep luminosity profile \citep{hos94b} and
a large central density, as seen in several young merger remnants (R. P. van
der Marel et al., in preparation; see also figure 2 in  \citeauthor{marel00}
\citeyear{marel00}). The presence of a central supermassive black hole would
also induce a strong nuclear power-law cusp \citep*{yng80,quin95}. Yet if both
galaxies contain supermassive central black holes which merge, the resulting
black hole binary would act as a dynamical slingshot and eject stars from the
center, thus lowering the stellar density there \citep{quin97,milo01}.

For these reasons we are undertaking a high resolution {\it Hubble Space
Telescope (HST)} survey of the nuclear regions in a sequence of merging
galaxies. The questions we want to answer include the following:  1. What is
the morphology of the  ionized gas distribution around the nucleus? Is it
clumpy, diffuse and extended, ring-like, or  compact and disk-like? Does this
morphology depend on the interaction  stage? 2. How is the current star
formation, as revealed by the H$\alpha$ line emission, distributed with respect
to the young stellar populations as revealed by their blue color (and,
eventually, spectra)?   3. How do the nuclear starbursts affect  the radial
color gradients of the merger remnants? The spatial resolution of the {\it HST}
(10--50 pc in nearby systems) is required to investigate these questions, which
are the focus of the current paper. In future papers, we will study the stellar
populations, kinematics, and evolution of the merger-induced starbursts of the
Toomre Sequence nuclei, using {\it HST} STIS spectra and NICMOS imaging.

\section{GALAXY SAMPLE AND OBSERVATIONS}
\label{s:sampleobs}
\subsection{Sample}
\label{ss:sample}

The Toomre Sequence \citep{tmr77} is a sample of 11 relatively nearby (within
$\sim$120\,Mpc) interacting and merging disk galaxies, which have been
arranged into a sequence according to the {\it putative} time before or since
merging. These systems were chosen because they exhibit conspicuous tidal
tails and ``main bodies that are nearly in contact or perhaps not even
separable'' \citep{tmr77}.  They span a range of dynamical phases, from
galaxies early in the merging process (e.g., NGC\,4038/39 and NGC\,4676) to
late-stage merger remnants (e.g., NGC\,3921 and NGC\,7252; see
Figure~\ref{f:mosaic} and Table~\ref{t:sample}). We want to emphasize the
word ``putative'', since the placing of a system in this sequence was  based
on the apparent degree of coalescence of the progenitor main bodies in
low-resolution optical photographical plates, assuming that the progenitors
resembled normal disk galaxies in the pre-encounter stage.

Because of the requirement of well-developed tidal tails, these systems have
suffered, or are suffering, encounters that are most likely of prograde sense
(with the disks rotating in the same sense as they orbit each other). In such a
scenario, the orbits of interacting galaxies decay as the galaxies lose orbital
energy and angular momentum via dynamical friction with their dark halos.
Ultimately, the galaxies coalesce and form a single merger remnant. On
these grounds we can call the Toomre Sequence an evolutionary sequence.
Because  it is an {\it optically} selected sample of merging galaxies, it
suffers less from dust obscuration than infrared-luminous samples, allowing the
nuclei to be studied at optical wavelengths. The Toomre Sequence has been
widely investigated by  ground-based observations
\citep{jos85,stan91,hib96,sch98,yun01,geo00},  space-based observations
\citep{fab97,read98,awa02,zez02a,zez02b}, and numerical  studies 
\citep{tmr72,bar88,stan91b,hos93,hib95b,bar98}. 

\begin{figure*}
\centering
\includegraphics[width=4in,angle=270]{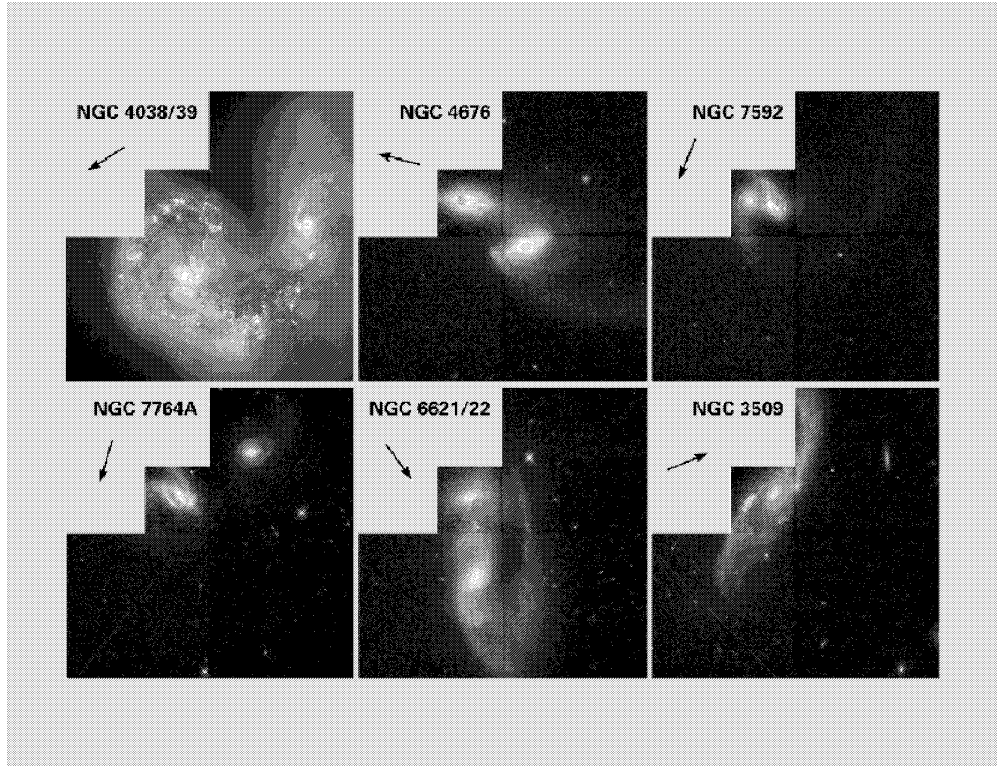}
\caption{Color composite, mosaiced F555W- and F814W-band 
images of the Toomre Sequence of merging galaxies, ordered from top left to
bottom right in the original order of \citet{tmr77}. The arrow in each
sub-panel indicates the North direction. The visible trails in the NGC\,7764A
and NGC\,3509 images are most probably satellite tracks.\label{f:mosaic}}
\end{figure*}

\begin{table*}
\caption{The Toomre Sequence.\label{t:sample}}
{\tiny
\begin{tabular}{ccccccccc}
\tableline
\tableline
Sequence Number & Galaxy & R.A. (J2000.0)\tablenotemark{\it a} & 
Dec. (J2000.0)\tablenotemark{\it a} & $cz$\tablenotemark{\it b} & 
Dist.\tablenotemark{\it c} & $\theta_{nuc}$\tablenotemark{\it d} & 
0\farcs1\tablenotemark{\it e} & Dynamical \\
& & (hh mm ss.ss) & (dd\degr mm\arcmin ss\farcs s) & (km~s$^{-1}$) & 
(Mpc) & (\arcsec) & (pc) & Model\tablenotemark{\it f} \\
\tableline
1. & NGC\,4038 & 12 01 53.06 & $-$18 52 01.3 & 1616 & 21.6 & 61.4 & 10  & 
Y\tablenotemark{\it g}\\
1. & NGC\,4039 & 12 01 53.54 & $-$18 53 09.3 & 1624 & 21.6 & 61.4 & 10  & 
Y\tablenotemark{\it g}\\
2. & NGC\,4676 NUC1 & 12 46 10.06 & +30 43 55.5 & 6613 & 88.2 & 37.1 & 43 & 
Y\tablenotemark{\it h}\\
2. & NGC\,4676 NUC2 & 12 46 11.17 & +30 43 21.2 & 6613 & 88.2 & 37.1 & 43 & 
Y\tablenotemark{\it h}\\
3. & NGC\,7592 NUC1 & 23 18 21.73 & $-$04 24 56.7 & 7280 & 97.1 & 13.0 & 47 & N\\
3. & NGC\,7592 NUC2 & 23 18 22.60 & $-$04 24 57.3\tablenotemark{\it i} & 7280 & 97.1 & 13.0 & 47 & N\\
4.& NGC\,7764A & 23 53 23.74 & $-$40 48 26.3 & 9162 & 122.2 & 0 & 59  & N\\
5. & NGC\,6621 & 18 12 55.25 & +68 21 48.5 & 6191 & 84.4 & 41.7 & 41 & N\\
5. & NGC\,6622 & 18 12 59.74 & +68 21 15.1 & 6466 & 84.4 & 41.7 & 41 & N\\
6. & NGC\,3509 & 11 04 23.59 & +04 49 42.4 & 7704 & 102.7 & 0 & 50  & N	\\
7. & NGC\,520 NUC1 & 01 24 34.89 & +03 47 29.9 & 2281 & 30.4 & 40.3 & 15 & 
Y\tablenotemark{\it j}\\
7. & NGC\,520 NUC2 & 01 24 33.30 & +03 48 02.4 & 2281 & 30.4 & 40.3 & 15 & 
Y\tablenotemark{\it j}\\
8. & NGC\,2623 & 08 38 24.11 & +25 45 16.6 & 5535 & 73.8 &  0 & 36 & N\\
9. & NGC\,3256 & 10 27 51.17 & $-$43 54 16.1 & 2738 & 36.5 & 0 & 18  & N\\
10. & NGC\,3921 & 11 51 06.96 & +55 04 43.1 & 5838 & 77.8 & 0 & 38 & N\\
11. & NGC\,7252 & 22 20 44.78 & $-$24 40 41.8 & 4688 & 62.5 & 0 & 30 & Y
\tablenotemark{\it k}\\
\tableline
\end{tabular} 
}
\tablenotetext{a}{Coordinates refer to the measured positions of the nuclei 
in our WFPC\,2 images, except for Nuc 1 in NGC\,520 where it was
measured from a ground-based $K$-band image (no clear nuclear position could be
determined in our {\it HST} images).}
\tablenotetext{b}{Heliocentric velocity, data taken from the NASA 
Extragalactic Database (NED), except for NGC\,4676, for which the value was taken 
from the Lyon Extragalactic Database (LEDA).}
\tablenotetext{c}{Distance for $\rm{H_0 = 75\,km\,s^{-1}\,Mpc^{-1}}$. In cases
where the two components have different systemic velocities, the distance to
both is the average of the two Hubble law distances.}
\tablenotetext{d}{Nuclear separation in arcseconds, calculated from the listed 
positions.}
\tablenotetext{e}{Spatial scale in pc corresponding to 0\farcs1.}
\tablenotetext{f}{Dynamical model means that an N-body simulation that tries
to reproduce the observed structure and kinematics of the merger exists.}
\tablenotetext{g}{Barnes 1988.} 
\tablenotetext{h}{Barnes 1998.}
\tablenotetext{i}{The position of Nuc 2 in NGC 7592 is highly uncertain, as
it is very difficult to determine the location of the nucleus in the optical
images.}
\tablenotetext{j}{Stanford \& Balcells 1991.}
\tablenotetext{k}{Hibbard \& Mihos 1995.}
\end{table*}

\subsection{Observations and Data Reduction}
\label{ss:obsreduc}

All images were taken with the WFPC2 camera onboard the {\it HST}. We used the
F555W and F814W filters, which mimic the better-known $V$- and $I$-bands, with
integration times of 320 seconds in each band, split into two exposures of 160
seconds to allow for cosmic ray rejection (see Table~\ref{t:expo} for more
information on the exposure times). Images in these bands for NGC\,4038/39
\citep{whit99}, NGC\,3921 \citep{schw96}, and NGC\,7252 \citep{mil97} already
existed in the {\it HST} archive. We used those images in our current study.
Narrow-band images covering the H$\alpha$+[\ion{N}{2}] lines were taken with
the F673N narrow-band filter in cases where the line emission from the target
fell within the wavelength range covered by this filter (NGC\,3509; NGC 7592),
or otherwise with the Linear Ramp Filter (LRF). The LRFs have a bandwidth of
about 1.3\% of the central wavelength. The position of the galaxy on the CCD
chip depends on the central wavelength and limits the field of view to about
$10\arcsec$~$\times$~$10\arcsec$ for the LRF.  The list of the adopted filters,
central wavelengths, and FWHM values of the filters together with integration
times are given in Table~\ref{t:expo}.  The Antennae, NGC\,4038/39, was imaged
earlier in the {\it HST} F658N H$\alpha$ filter by \citet{whit99}, and we used
these data in our work.

\setcounter{figure}{0}
\begin{figure*}
\centering
\includegraphics[width=4in,angle=270]{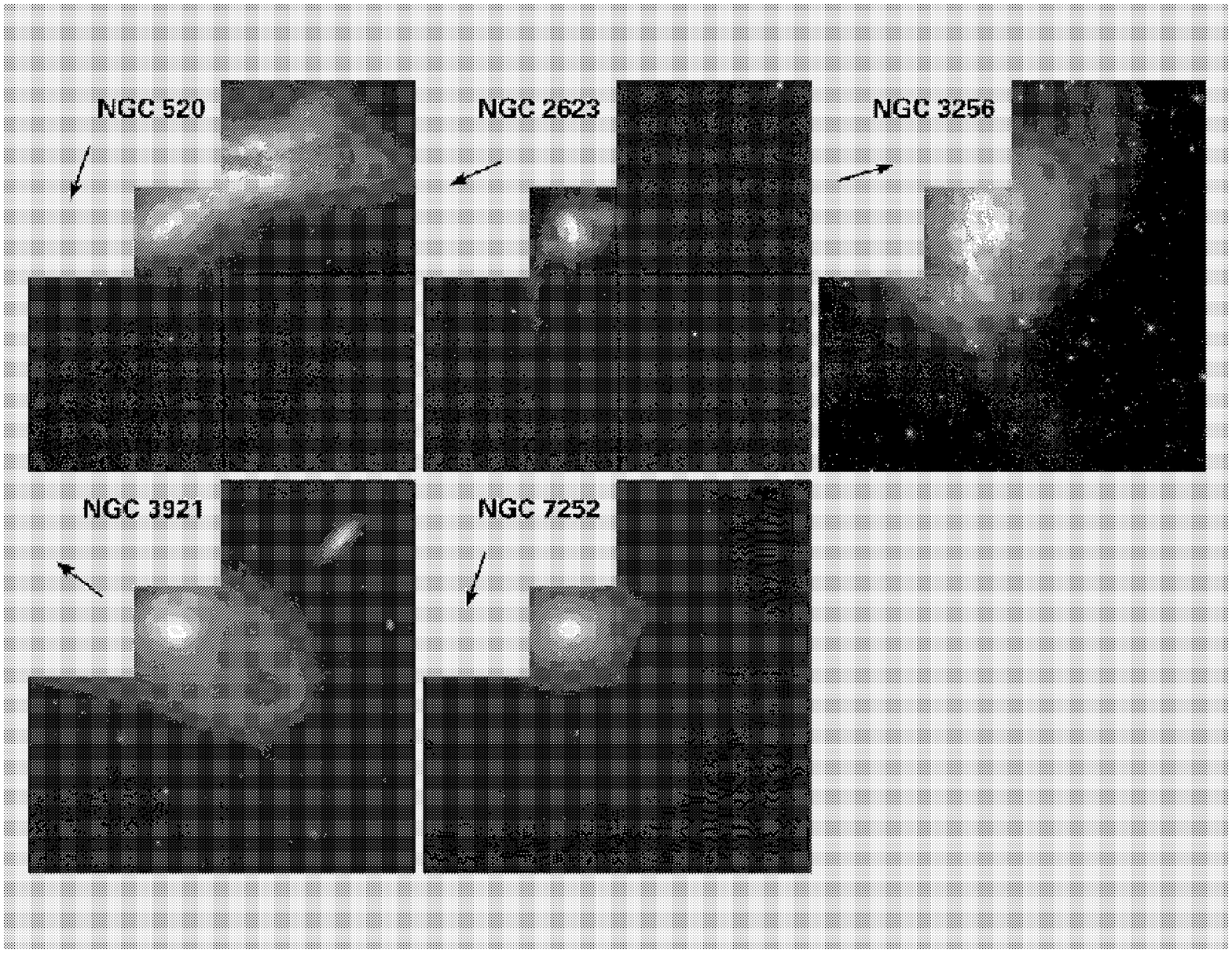}
\caption{Continued.}
\end{figure*} 

We employed the STSDAS task WFIXUP to interpolate (in the x-direction) over bad
pixels as identified in the data quality files. We also used the STSDAS task
WARMPIX to correct consistently warm pixels in the data, using the most recent
warm pixel tables. The STSDAS task CRREJ was used to combine the two 160 second
exposures. This step corrects most of the pixels affected by cosmic rays in the
combined image. In general, a few cosmic rays remain uncorrected, mostly when
the same pixel was hit in both exposures. Also, a small number of hot pixels
remain uncorrected because they are not listed even in the most recent warm
pixel tables. We corrected these with the IRAF task COSMICRAYS, setting the
``threshold'' and ``fluxratio'' parameters to  values selected  by a careful
comparison of the images before and after correction, to ensure that only
questionable pixels were replaced.  The rest of the reduction was done with the
standard WFPC2 pipeline tasks using the best reference files available. The LRF
exposures were flat-fielded using the narrow-band F656N and F673N (whichever
was closer in wavelength) flat fields.  We also created mosaiced F555W and
F814W images with the IRAF/STSDAS task WMOSAIC. Color composite mosaics of
these images, which show the environment around the nuclei, and tidal tails in
some systems, are displayed in Figure~\ref{f:mosaic}.

The photometric calibration, and conversion to Johnson $V$- and $I$-bands was
performed according to guidelines in \cite{hol95}.  A $K$-correction has not
been applied to any of our measurements. Since the throughput of the
narrow-band filters does not vary much at the wavelengths of the lines, it was
possible to calibrate the line fluxes by assuming zero width for the lines
(monochromatic). We used the IRAF/STSDAS SYNPHOT task BANDPAR to compute the
conversion from counts~sec$^{-1}$ to ergs~sec$^{-1}$~cm$^{-2}$. The results
were the same within errors to those produced by the SYNPHOT task CALCPAR.

Using the known relative orientation of the various chips on WFPC2, we rotated
the chips which contained the line emission and the PC-chip $V$ and $I$ images
to the same orientation. We used stars in the images to  register the frames. 
The final emission-line images were constructed by accounting for the different
scales of images on various chips (usually the PC chip and WF2 chip), and
scaling down the $V$- and $I$-band images to the expected count levels
corresponding to the wavelength and band of the line image, with the help of
the WFPC2 exposure time calculator. We then combined the broad-band images by
taking their geometric mean and performed a final adjustment to this combined
continuum image by selecting areas well outside emission-line regions (with
pure stellar emission) and comparing fluxes in corresponding areas in the
continuum and line images. Finally, we subtracted the scaled continuum image
from the line image. Using the combination of $V$ and $I$ images eliminates the
extinction terms in  the case of foreground extinction, as explained in detail
by \citet{ver99}. In some cases, the residual image shows negative pixels,
which are most likely due to strong color gradients near the nuclei. We
estimate that the final uncertainty in the line fluxes is no better than 50\%,
based on varying the scaling of the continuum image within acceptable limits
and comparing the derived total line fluxes.

\begin{table*}
\caption{WFPC2 EXPOSURES WITH THE CORRESPONDING CHIP ON WHICH THE NUCLEI ARE 
CENTERED, AND INTEGRATION TIMES.\label{t:expo}}
\tiny{
\begin{tabular}{lccccc}
\tableline
\tableline
Object & Chip & Filter & $\lambda$$_{0}$\tablenotemark{a} & $\Delta$$\lambda$\tablenotemark{b} & $t$(sec)\tablenotemark{c} \\
\tableline
NGC\,4038/39\tablenotemark{d} & WF2/WF4 & F555W & 5407 & 1223 & 4 $\times$ 1100\\
NGC\,4038/39\tablenotemark{d} & WF2/WF4 & F814W & 5407 & 1223 & 4 $\times$ 500\\ 
NGC\,4038/39\tablenotemark{d} & WF2/WF4 & F658N & 6591 & 28.5 & 3800 \\ 
NGC\,4676 Nuc 1 & PC1 & F555W & 5407 & 1223 & 2 $\times$ 160 \\
NGC\,4676 Nuc 1 & PC1 & F814W & 7940 & 1758 & 2 $\times$ 160\\ 
NGC\,4676 Nuc 1 & WF2 & LRF & 6715.3 & 80.1 & 2 $\times$ 600 \\
NGC\,4676 Nuc 2 & PC1 & F555W & 5407 & 1223 & 2 $\times$ 160\\ 
NGC\,4676 Nuc 2 & PC1 & F814W & 7940 & 1758 & 2 $\times$ 160\\ 
NGC\,4676 Nuc 2 & WF2 & LRF & 6714.4 & 79.9 & 2 $\times$ 600\\ 
NGC\,7592 & PC1 & F555W & 5407 & 1223 & 2 $\times$ 160 \\ 
NGC\,7592 & PC1 & F814W & 7940 & 1758 & 2 $\times$ 160 \\ 
NGC\,7592 & WF2 & F673N & 6732 & 47.2 & 2 $\times$ 600 \\ 
NGC\,7764A & PC1 & F555W & 5407 & 1223 & 2 $\times$ 160 \\
NGC\,7764A & PC1 & F814W & 7940 & 1758 & 2 $\times$ 160 \\
NGC\,7764A & WF2 & LRF & 6770.8 & 84.2 & 2 $\times$ 600 \\
NGC\,6621 & PC1 & F555W & 5407 & 1223 & 2 $\times$ 160 \\ 
NGC\,6621 & PC1 & F814W & 7940 & 1758 & 2 $\times$ 160\\  
NGC\,6621 & WF2 & LRF & 6706.2 & 79.3 & 2 $\times$ 700\\  
NGC\,6622 & PC1 & F555W & 5407 & 1223 & 2 $\times$ 160\\  
NGC\,6622 & PC1 & F814W & 7940 & 1758 & 2 $\times$ 160\\  
NGC\,6622 & WF2 & LRF & 6711.5 & 79.9 & 2 $\times$ 700\\  
NGC\,3509 & PC1 & F555W & 5407 & 1223 & 2 $\times$ 160\\  
NGC\,3509 & PC1 & F814W & 7940 & 1758 & 2 $\times$ 160\\  
NGC\,3509 & PC1 & F673N & 6732 & 47.2 & 2 $\times$ 600\\  
NGC\,520 Nuc 1 & PC1 & F555W & 5407 & 1223 & 2 $\times$ 160 \\
NGC\,520 Nuc 1 & PC1 & F814W & 7940 & 1758 & 2 $\times$ 160 \\ 
NGC\,520 Nuc 1 & PC1 & LRF & 6619.2 & 75.5 & 2 $\times$ 600 \\
NGC\,520 Nuc 2 & PC1 & F555W & 5407 & 1223 & 2 $\times$ 160 \\
NGC\,520 Nuc 2 & PC1 & F814W & 7940 & 1758 & 2 $\times$ 160 \\
NGC\,520 Nuc 2 & PC1 & LRF & 6617.1 & 75.8 & 2 $\times$ 600 \\
NGC\,2623 & PC1 & F555W & 5407 & 1223 & 2 $\times$ 160 \\     
NGC\,2623 & PC1 & F814W & 7940 & 1758 & 2 $\times$ 160 \\     
NGC\,2623 & WF2 & LRF & 6691.4 & 78.6 & 2 $\times$ 600 \\ 
NGC\,3256 Nuc 1 & PC1 & F555W & 5407 & 1223 & 2 $\times$ 160 \\
NGC\,3256 Nuc 1 & PC1 & F814W & 7940 & 1758 & 2 $\times$ 160 \\
NGC\,3256 Nuc 1 & WF2 & LRF & 6631.4 & 75.9 & 2 $\times$ 600 \\
NGC\,3256 Nuc 2 & PC1 & F555W & 5407 & 1223 & 2 $\times$ 160 \\
NGC\,3256 Nuc 2 & PC1 & F814W & 7940 & 1758 & 2 $\times$ 160 \\
NGC\,3256 Nuc 2 & WF2 & LRF & 6631.4 & 75.9 & 2 $\times$ 600 \\
NGC\,3921\tablenotemark{e} & PC1 & F555W & 5407 & 1223 & 2 $\times$ 1200 \\
NGC\,3921\tablenotemark{e} & PC1 & F814W & 7940 & 1758 & 2 $\times$ 900 \\ 
NGC\,3921 & WF2 & LRF & 6698.9 & 79.0 & 2 $\times$ 700 + 800 \\
NGC\,7252\tablenotemark{f} & PC1 & F555W & 5407 & 1223 & 3600\\ 	   
NGC\,7252\tablenotemark{f} & PC1 & F814W & 7940 & 1758 & 2400\\ 	   
NGC\,7252 & WF2 & LRF & 6673.9 & 77.9 & 2 $\times$ 700 + 600\\
\tableline
\end{tabular}
}
\tablenotetext{a}{Central wavelength in \AA~[from the WFPC2 Handbook, Table 3.1 
\citet{bir20}, for the standard filters; as specified in the observations for the 
LRF filters].}
\tablenotetext{b}{Filter width in \AA~[from the WFPC2 Handbook, Table 3.1
\citet{bir20}, for the standard filters; equivalent Gaussian FWHM from the IRAF 
CALCPHOT task for the LRF filters].}
\tablenotetext{c}{Integration time.}						
\tablenotetext{d}{Data from Whitmore et al. 1999.}
\tablenotetext{e}{Data from Schweizer et al. 1996.}	
\tablenotetext{f}{Data from Miller et al. 1997.}
\end{table*}

Color index images were created by dividing the F555W image by the F814W image,
taking the logarithm of the result, and correcting for differences in the color
terms, using the synthetic calibration from \citet{hol95}.  For our new imagery
(i.e., all systems except NGC 4038/39, NGC 3921, and NGC 7252), many regions in
the raw color index images are of low  signal-to-noise. We therefore
constructed smoothed color index maps  using an adaptive filtering procedure,
as described by \citet{sco00},  where areas with lower signal-to-noise ratios
were smoothed by a  larger boxcar. These images were then masked so as to only
show areas with adequate signal-to-noise (signal-to-noise greater than two in 
both $V$- and $I$-images after being smoothed with an 11$\times$11  pixel
boxcar median filter). The resulting smoothed color index images, together with
the unsmoothed broad-band images and H$\alpha$+[\ion{N}{2}] images, are shown
in Figures~\ref{f:n4038}--\ref{f:n7252}.

\section{GALAXY MORPHOLOGY}
\label{s:images}

In the following we give a brief description of each system, based on
Figures~\ref{f:n4038}--\ref{f:n7252}, together with a few references to earlier
work. We use a Hubble constant of $\rm{H_0}=75\,\rm{km\,s^{-1}\,Mpc^{-1}}$
throughout this paper.

\subsection{NGC\,4038 and NGC\,4039 (The Antennae)}
\label{ss:ngc4038_9}

Perhaps the best-known system in the sequence is the Antennae.  We use the
spectacular {\it HST} $V, I$, and H$\alpha$ images of  this system from
\citeauthor{whit99} (\citeyear{whit99}; see also \citealt{whit95} and
Figure~\ref{f:mosaic}). These images show the chaotic dust lanes and prevalent
star formation in the interface between the two galaxies. They also reveal the
redder color of the underlying population in the bulge component of each
galaxy.  The focus of the \citeauthor{whit99} papers was on the abundant young
star clusters which have apparently formed as the result of the interaction of
the two galaxies. The properties and evolution of such clusters in a merger
event have been discussed extensively in a series of papers by Whitmore and his
collaborators \citep*[e.g.,][]{whit95,schw96,mil97,whit97,whit99,zhang01}. The
nuclear regions of merging galaxies are potential formation sites of new super
star clusters \citep{bekki01}. We instead investigate the morphology and star
formation activity in the central few hundred parsecs.

\begin{figure*}[th]
\centering
\includegraphics[width=6in,angle=270]{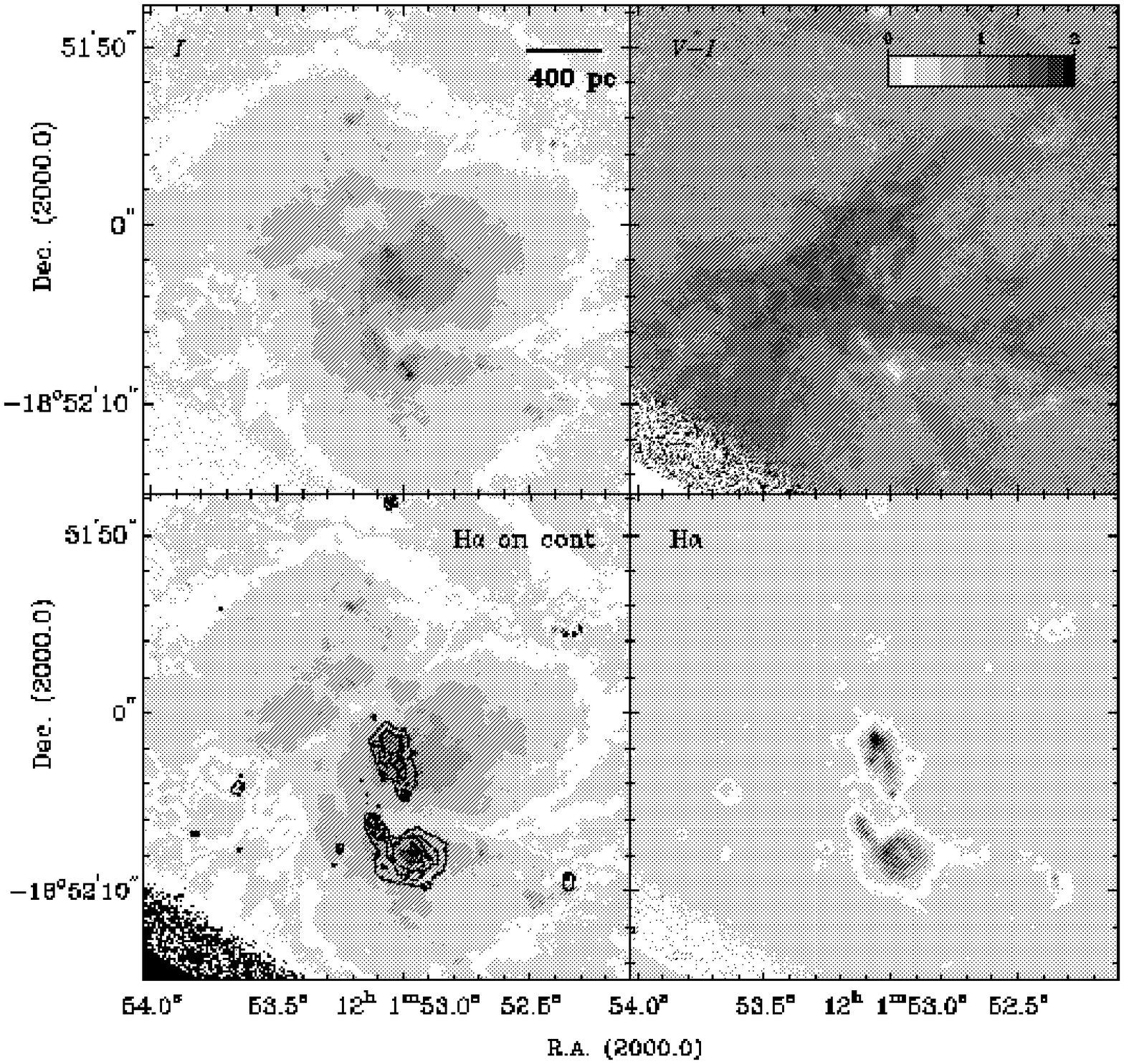}
\caption{NGC\,4038. Upper left: $I$ image. Upper right: $V-I$ color index image. 
Lower left: continuum (gray-scale) and H$\alpha$ + [\ion{N}{2}] (contours) image. 
Lower right: H$\alpha$ gray-scale image. The gray-scale levels are arbitrary and 
were adjusted to show details in the nuclear region. The H$\alpha$ contour levels 
in the lower left image are at (10\%, 30\%, 50\%) of 7.2$\times$10$^{-17}$ 
ergs~s$^{-1}$~cm$^{-2}$ per pixel. North is up and east is to the left in this and 
in the subsequent galaxy images.\label{f:n4038}}
\end{figure*}

The nuclear region of the northern component of the Antennae, NGC\,4038, is
bracketed by star clusters in the $V$ and $I$ images. The position of the 
nucleus is most likely near the knot "J" in the classification scheme of 
\citet{whit95}, and close to the nuclear position identified in the radio 
continuum by \citeauthor{zhang01} (\citeyear{zhang01}; using the radio
continuum data of \citeauthor{neff00} \citeyear{neff00}) and in the
near-infrared by  \citet{meng01}. In our I-band image, this is located near the
center of the frame displayed in Fig.~\ref{f:n4038}. There we see an elongated
ring-like structure, with an axis ratio of about 0.3 and a diameter of about
2\farcs 5~(260 pc), pointing north-northeast. The coordinates of the optical
knot that we identified as the nucleus of NGC\,4038 in the WFPC2 images are
given in Table~\ref{t:sample}. They agree with the radio continuum position to
within one arcsecond and they are within about 3 arcseconds of the estimated
near-infrared position of the nucleus. The absolute accuracy of the coordinates
reported in this paper is $\sim$1\arcsec. This is the intrinsic accuracy of the
{\it HST} guide star coordinate system. The H$\alpha$ image (see
Fig.~\ref{f:n4038}) also shows the elongated ring structure with a peak in the
north-northeastern corner at the position of the star clusters seen in the $V$-
and $I$-band images. The $V-I$ colors of the clusters in the ring-like
structure vary from 0.4 to 1.0, while a typical color outside and inside the
ring is 1.5, reaching up to 1.9 in the dust lanes. Most of the dust is
concentrated on the southeastern side of the elongated ring-like structure.
Near-infrared high resolution {\it HST} NICMOS observations of the nucleus will
be reported in a future paper. They promise to shed more light on the issue of
the true location of the nucleus in NGC\,4038. Previous ground-based
near-infrared imaging  \citep*{bus90,hib01,meng01} does not have high enough spatial
resolution to indicate the exact location  of the nucleus.

\begin{figure*}[th]
\centering
\includegraphics[width=6in,angle=270]{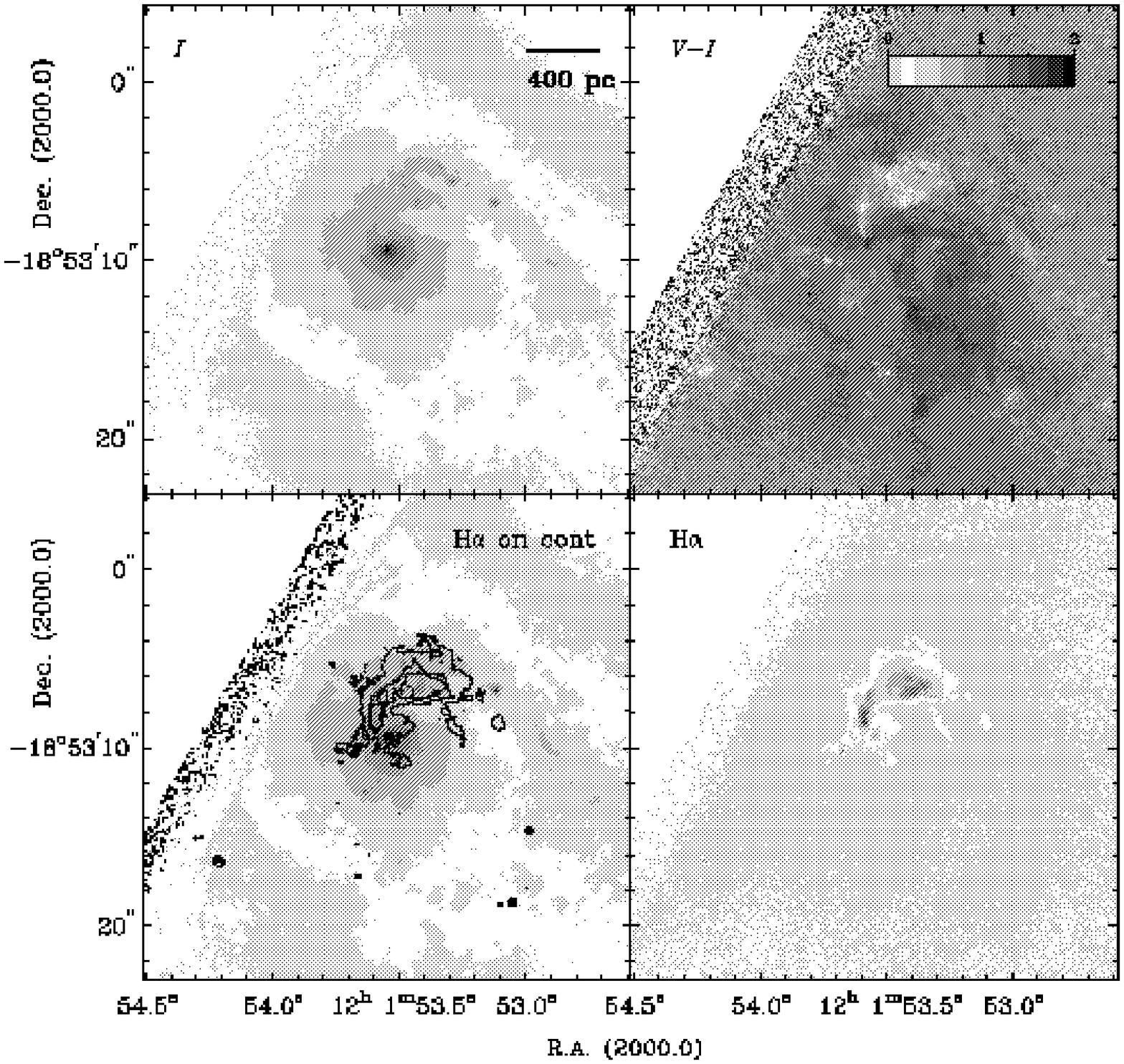}
\caption{NGC\,4039, as in Figure~\ref{f:n4038}. 
The H$\alpha$ contour levels in the lower left image are at (10\%, 30\%, 50\%) of 
7.5$\times$10$^{-17}$ ergs~s$^{-1}$~cm$^{-2}$ per pixel.\label{f:n4039}}
\end{figure*}

The nucleus of NGC\,4039, the southern component of the Antennae, is much
easier to recognize in the $V$ and $I$ images (see Fig.~\ref{f:n4039}). Our
identified nuclear position (the position of the optical peak) in
Table~\ref{t:sample} agrees to within about two arcseconds of the estimated
nuclear positions from the radio continuum data \citep{zhang01} and
near-infrared data \citep{meng01}. The nucleus is surrounded by dust patches
and dust lanes. There is also a minor peak of H$\alpha$ emission at the
location of the nucleus, which could be due to continuum subtraction
uncertainties. Much more intense H$\alpha$ emission is seen in the arm
connecting to the nucleus. The $V-I$ color of the nucleus of NGC\,4039 is about
1.0, and the surrounding area has a color around 1.3, with values of 1.7 in a
dusty region close to the nucleus. 

Neither NGC\,4038 nor NGC\,4039 is classified as a Seyfert or LINER nucleus.
These nuclei are instead hosts of mild starbursts
\citep[e.g.,][]{daha85,ver86}. Further evidence for the starburst nature of
these nuclei comes from X-ray observations with the ROSAT High-Resolution
Imager \citep{fab97} and with Chandra \citep{zez02a,zez02b}. The northern X-ray
nucleus (NGC\,4038) has a soft spectrum, which  hints at thermal emission and
is probably related to a hot wind, whereas the thermal + power-law spectrum of
the southern X-ray nucleus (NGC\,4039) indicates a hot ISM and a contribution
from X-ray binaries. 

\subsection{NGC\,4676 (The Mice)}
\label{ss:ngc4676}

\begin{figure*}[th]
\centering
\includegraphics[width=6in,angle=270]{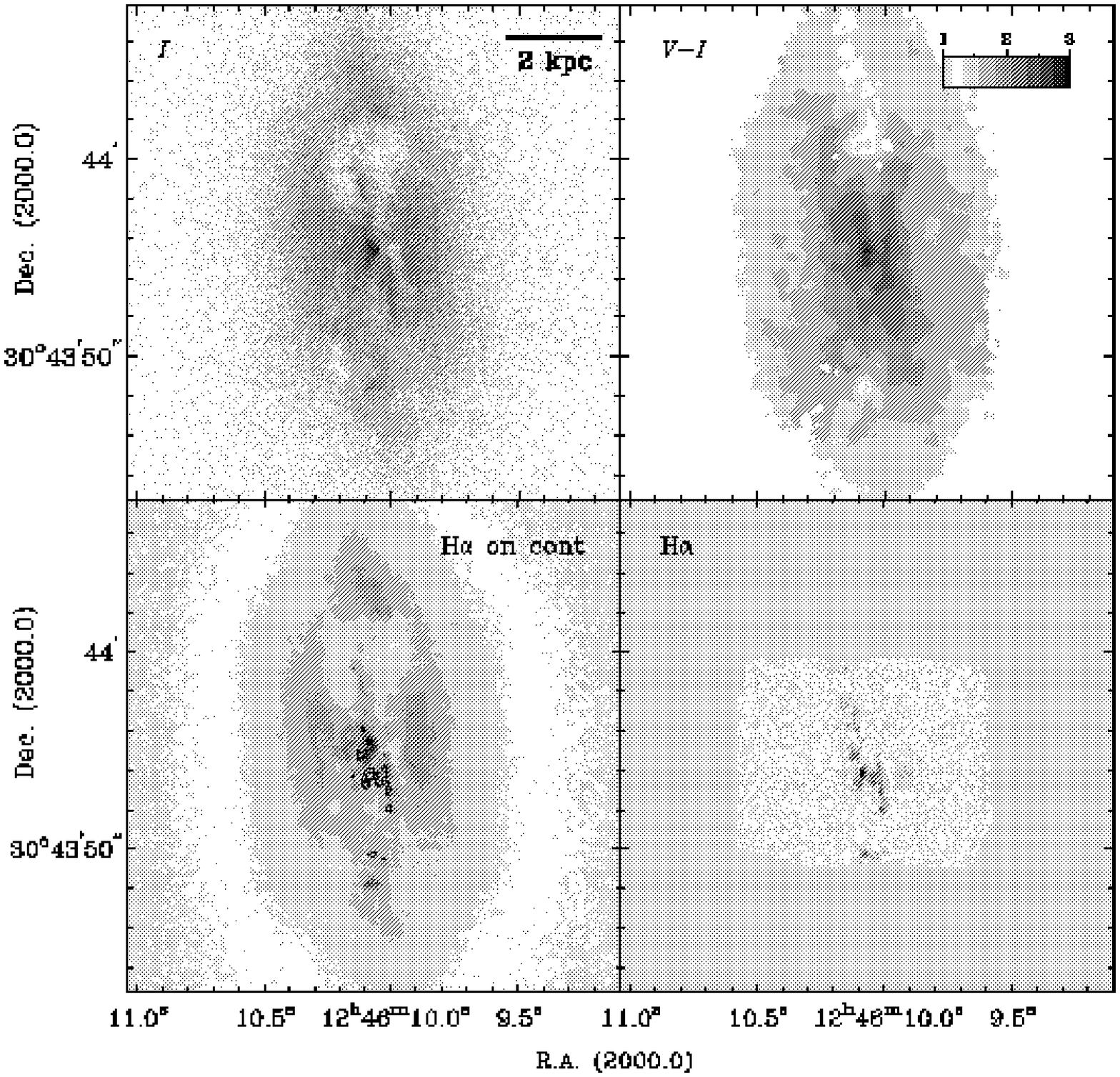}
\caption{NGC\,4676A or Nuc 1, as in Figure~\ref{f:n4038}. 
This color index image and many of the subsequent color index images only
show areas where the signal-to-noise ratio was sufficiently high to reveal
a meaningful, smooth color. The H$\alpha$ contour levels in the lower left image 
are at (20\%, 90\%) of 1.2$\times$10$^{-16}$ ergs~s$^{-1}$~cm$^{-2}$ per pixel.
\label{f:n4676n1}}
\end{figure*}

The Mice is a pair of spiral galaxies with only moderately active nuclei
(classified as LINER-type by \citealt{keel85}). The nucleus of NGC\,4676A or
Nuc~1 (the northern component of the pair) is covered by dust in our optical
$V$ and $I$ images (Fig.~\ref{f:n4676n1}). This dust is likely associated with
the dense edge-on molecular disk imaged  in CO(1-0) by \citet{yun01}. Adopting
standard conversion factors, the  observed peak CO flux density suggests that
the nucleus of NGC 4676A is  hidden beneath $A_V\sim$~60 magnitudes of
extinction. 

Just south of the area of heavy dust obscuration near the base of the
northern tail is a V-shaped structure (with the V being sideways and opening
to the east, and the tip of the V at R.A. 12$^{\rm h}$46$^{\rm m}$10\fs 1~and
Dec. $30\degr$43\arcmin55\farcs 0) of brighter emission from young stellar
clusters near the center of the main body.  About 7\arcsec~(3 kpc) to the
south one can see a triangle-shaped structure of young clusters.  The tip of
the central V-shaped cluster lies near the peak emission in a ground-based K'
image (J. Hibbard, unpublished; see also \citeauthor{bus92}
\citeyear{bus92}), so we assume that the nucleus lies near this position. The
H$\alpha$ emission from NGC\,4676A is weak. In general, the emission is
elongated in the north--south direction of the main body. The V-shaped
structure has very red $V-I$ colors above 2.2. Perhaps even more
surprisingly, the color of the heavily extinguished dust patch near the base
of the tail is actually {\it bluer} than its surroundings, having an average
$V-I$ value less than 1. We interpret this region as scattered light from
young stars which are mostly hidden by the dust. A hint of this  young star
population is seen in the $V$ image (not shown here). Such a young star
forming  region can be seen even more clearly in the new, deeper ACS images
of the  Mice \citep{for02}. In contrast to the V-shaped structure, the
triangle-shaped structure of clusters near the southern end of the main body
has bluer colors than its surroundings, with $V-I$ values around 1 or
slightly below it.

\begin{figure*}[th]
\centering
\includegraphics[width=6in,angle=270]{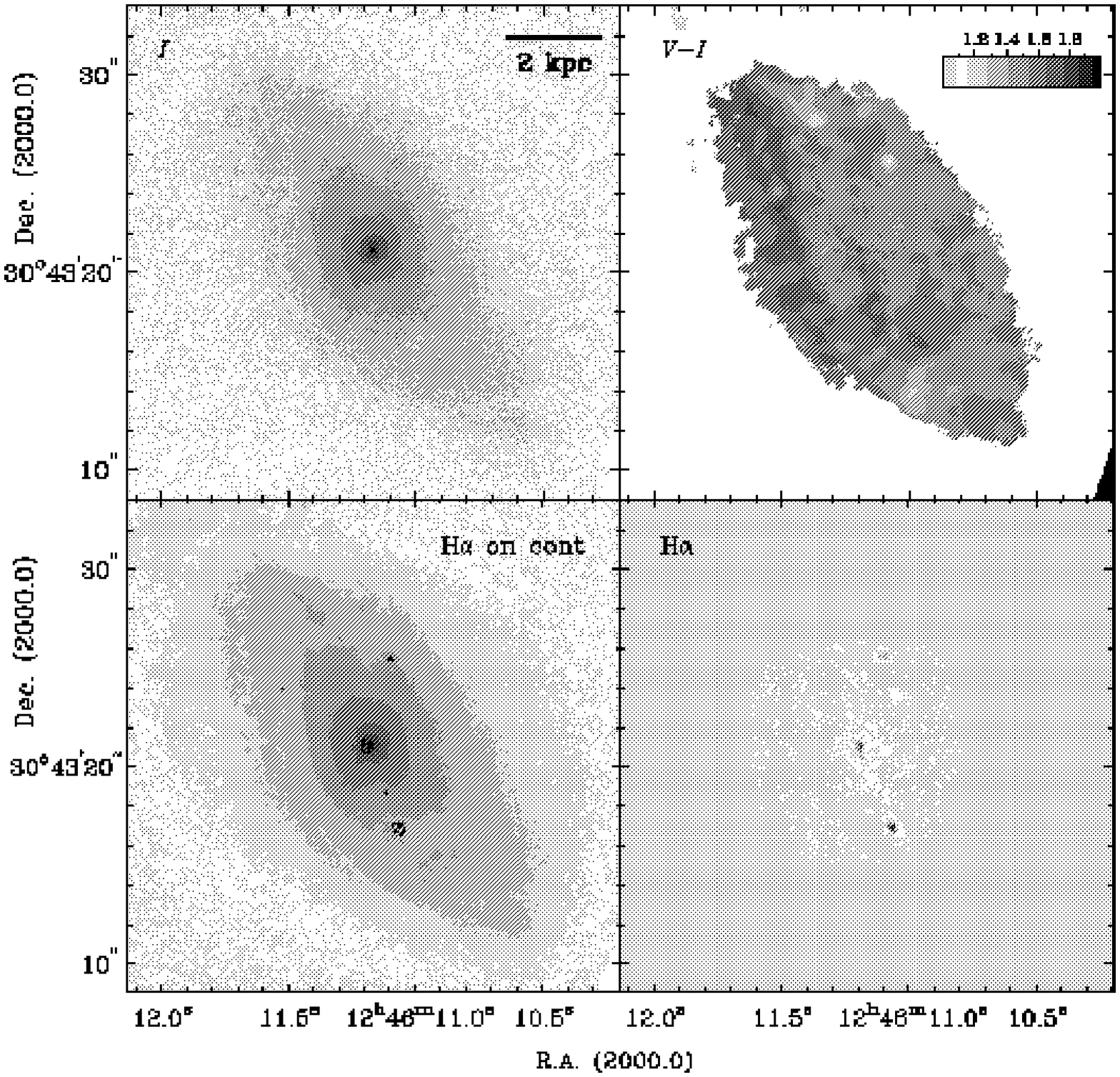}
\caption{NGC\,4676B or Nuc 2, as in Figure~\ref{f:n4038}. 
The H$\alpha$ contour levels in the lower left image are at 
(20\%, 50\%)  of 3.8$\times$10$^{-17}$ ergs~s$^{-1}$~cm$^{-2}$ per pixel.
\label{f:n4676n2}}
\end{figure*}

The nucleus is easily identified in NGC\,4676B or Nuc~2. It is surrounded by a
rather amorphous disk, and a dust lane that seems to wrap around the whole
disk, ending near the location of the nucleus (Fig.~\ref{f:n4676n2}). There is
a peak of H$\alpha$ emission offset by about 0\farcs 2 from the nucleus of
NGC\,4676B, but in general the H$\alpha$ emission is weak and patchy. The
nucleus has $V-I$ colors close to 1.3. A dust patch near the nucleus has a
$V-I$ color as red as 1.7. The majority of the disk outside the dust lanes and
dust patches has a $V-I$ color of 1.2--1.3. The overall appearance of
NGC\,4676B, particularly the well-defined bulge/nucleus, is consistent with it
being of an earlier Hubble-type than its interacting partner, NGC\,4676A
\citep{yun01}.

\subsection{NGC\,7592}
\label{ss:ngc7592}

\begin{figure*}[th]
\centering
\includegraphics[width=6in,angle=270]{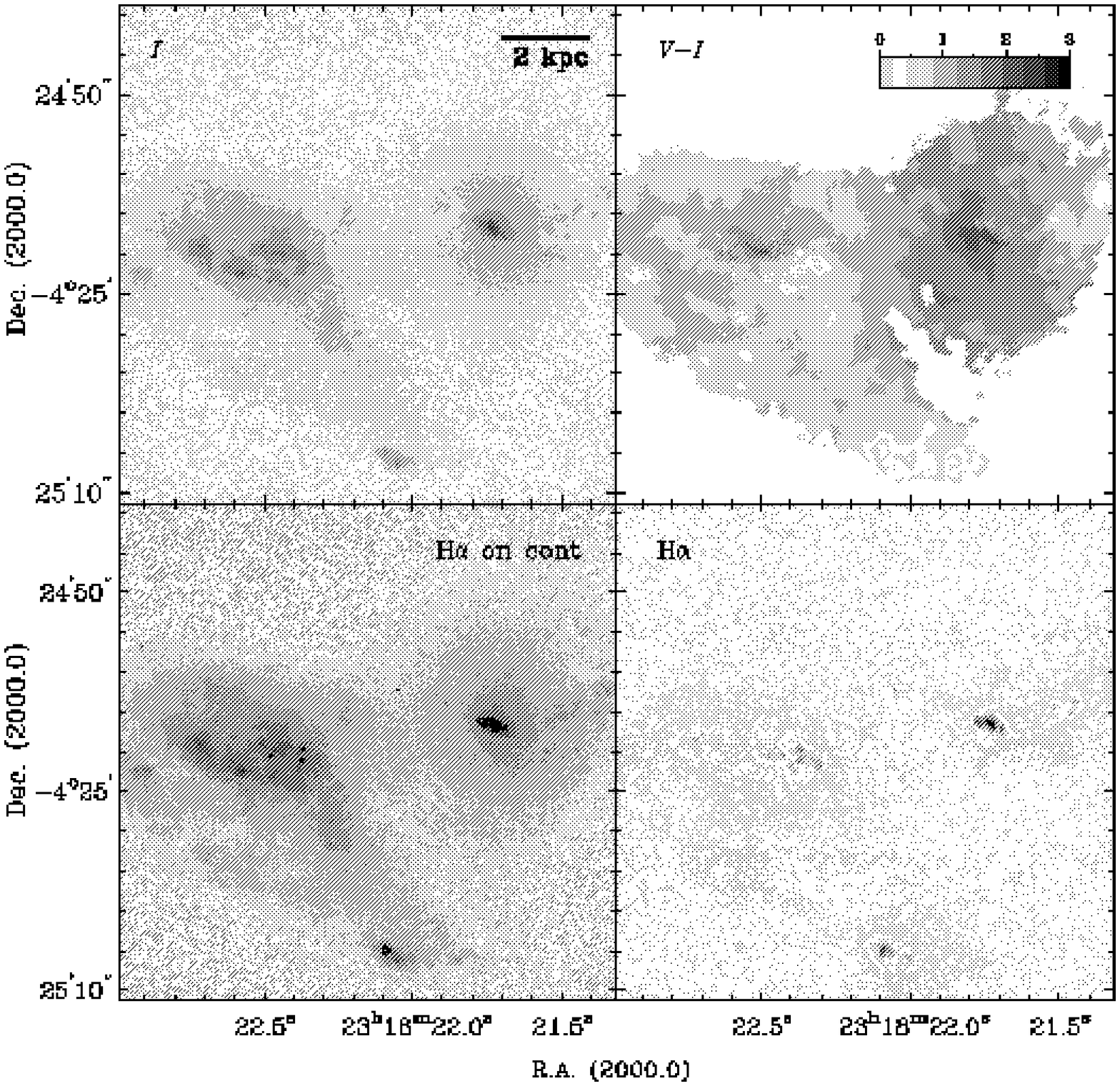}
\caption{NGC\,7592, as in Figure~\ref{f:n4038}. 
The H$\alpha$ contour levels in the lower left image are at 
(10\%, 30\%, 50\%) of 1.2$\times$10$^{-15}$ ergs~s$^{-1}$~cm$^{-2}$ per pixel.
\label{f:n7592}}
\end{figure*}

This is the third most distant system in the Toomre Sequence, and  the main
bodies of both galaxies fit within the PC chip of the WFPC2. NGC\,7592A or
Nuc~1, further to the west, has an active Seyfert  2 nucleus, whereas
NGC\,7592B or Nuc~2 is classified as a  starbursting system 
\citep*{daha85,raf92,kew01,hat02}.  NGC\,7592A has a much better defined
nucleus or bulge than NGC\,7592B, consistent with the suggestion of
\citet{hat02}, who argue that NGC\,7592A is of an earlier Hubble-type than
NGC\,7592B. \citet{bus90}  claim that NGC\,7592A is seen almost face-on,
because of its roundish appearance in ground-based optical and $K$-band images.
If so, this galaxy appears to possess a bar-like structure in the bulge,
although the bar classification is uncertain due to patchy emission and
one-sided dust. Similarly, \citet{bus90} argued that NGC\,7592B is more highly
inclined to the line-of-sight. This, together with its later Hubble-type,
explains why no clear nucleus is seen in this system. A comparison to the
ground-based $K$-band image of \citet{bus90} suggests that the nucleus of
NGC\,7592B lies near the center of the ring-like structure of bright clumps
seen in Figure~\ref{f:n7592}. We are not able to identify any specific clump
with the nucleus.

The H$\alpha$ image shows abundant emission around the nucleus and the ``bar''
of NGC\,7592A, and emission from the clumps in the ring of NGC\,7592B, with a
strong central depression in emission. The nucleus of NGC\,7592A has blue
colors with $V-I$ around 0.8, contrasting to a dust lane lying on one side of
the bar with a $V-I$ color close to 2.0. The southern side of the ring
structure in NGC\,7592B has very blue colors down to $V-I$ of 0.3, but there is
a region of red colors of $V-I$ up to 1.9 in the northern part of the ring.
Although there is dust visible at that location as well, the red color may
suggest that the underlying red nucleus lies at this location.

\subsection{NGC\,7764A}
\label{ss:ngc7764}

\begin{figure*}[th]
\centering
\includegraphics[width=6in,angle=270]{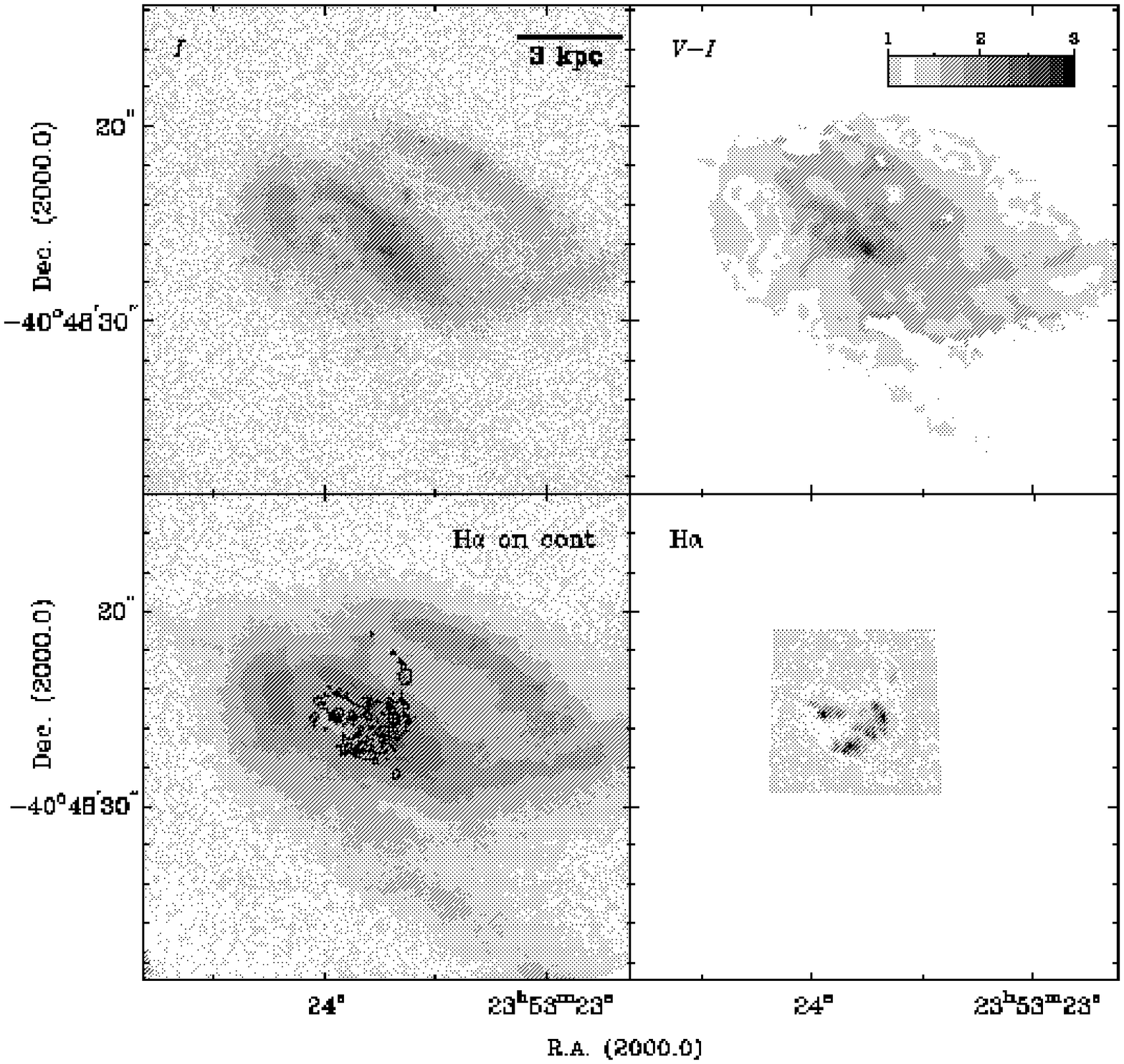}
\caption{NGC\,7764A, as in Figure~\ref{f:n4038}. 
The H$\alpha$ contour levels in the lower left image are at (10\%, 30\%, 50\%) of 
2.4$\times$10$^{-16}$ ergs~s$^{-1}$~cm$^{-2}$ per pixel.\label{f:n7764A}}
\end{figure*}

This is the most distant system in our sample at $D$~=~122 Mpc.  The location
of NGC\,7764A in the Toomre Sequence, among galaxies with clearly separate
nuclei (Fig.~\ref{f:mosaic}) suggests that Toomre thought the nuclei of the two
systems were still distinct.  However, our {\it HST} images
(Figure~\ref{f:n7764A}) do not obviously  support the presence of two distinct
nuclei. Future NICMOS near-infrared data will be critical in identifying the
likely remains of the interacting galaxies. For now we only identify one main
component from which both the tails are emanating. There is also a barred
spiral galaxy, presumably at a similar redshift, displaced about
40\arcsec~(23.7 kpc in projected distance) to the southeast of the main system
(imaged on one of our WF chips, see Fig.~\ref{f:mosaic}), and a third system 
with apparent tidal streamers displaced about 40\arcsec~to the  northwest, just
outside our PC image. 

Our PC image reveals a shred or linear feature  about 7\arcsec~(4.1 kpc) to the
northwest of the main body of NGC\,7764A. About a dozen bright, blue
($V-I\sim$~0.6--0.8) star clusters lie within this filament. Between this
filament and the center of Fig.~\ref{f:n7764A} is a very bright, very blue
($V-I\sim$~0.22) cluster. This cluster has associated H$\alpha$ emission with
a``head-tail'' morphology, with the cluster at the ``head'', and the ``tail''
pointing to the north-northeast.  Following the ``head'' towards the center of
the image in Fig.~\ref{f:n7764A}, there is a string of \ion{H}{2} regions,
crossed by dust lanes. The centralmost \ion{H}{2} region is weaker than the
regions to its northwest and southeast, and is associated with two adjacent 
bright optical clusters. These two clusters lie near the region of very red
colors  in the $V-I$ map, with the reddest region ($V-I\sim$2.4) associated
with the  eastern cluster. This strong dust concentration may indicate that
this is the  location of one of the nuclei. The western clump of the double 
cluster is bluer ($V-I\sim$~1.5) and brighter.

There are more bright H{\sc ii} regions about 3\arcsec~(1.8~kpc) to the
northeast of the double cluster, associated with another bright star cluster
embedded in dust. Further to the east of this is an amorphous luminous region
with rather blue colors ($V-I\sim$~0.6--0.8).

\subsection{NGC\,6621 and NGC\,6622}
\label{ss:ngc6621_2}

\begin{figure*}[th]
\centering
\includegraphics[width=6in,angle=270]{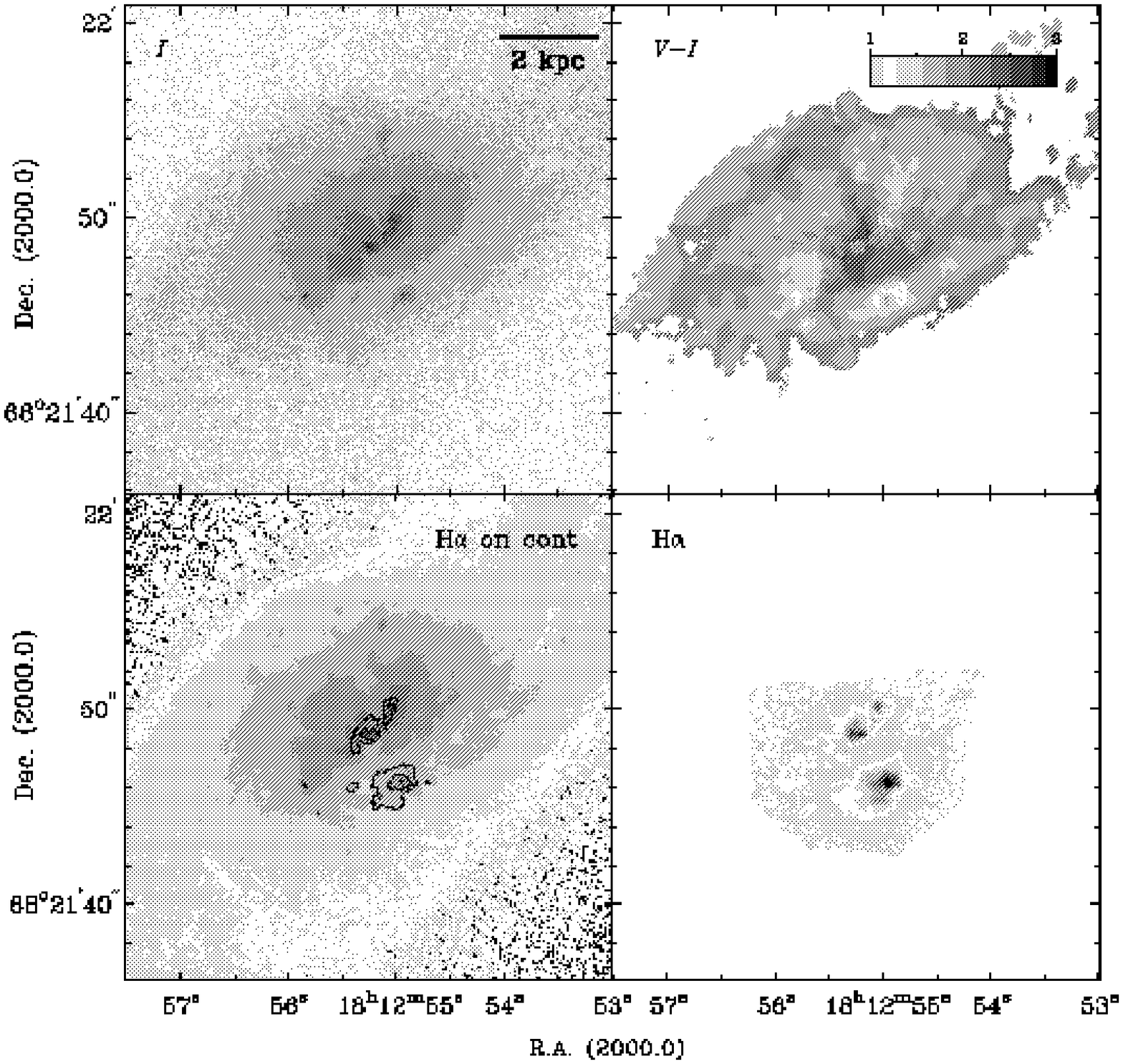}
\caption{NGC\,6621, as in Figure~\ref{f:n4038}. 
The H$\alpha$ contour levels in the lower left image are at (5\%, 20\%, 90\%) of 
1.9$\times$10$^{-15}$ ergs~s$^{-1}$~cm$^{-2}$ per pixel.\label{f:n6621}}
\end{figure*}

The northern component of this interacting galaxy pair, NGC\,6621, has a dusty
central region punctuated by patchy star clusters and ionized gas emission
(Fig.~\ref{f:n6621}).  Two major dust lanes lead into the central region. One
of them comes in from the northwest, following the long tidal tail at larger
radii. The other dust lane intersects the nuclear region from the north,
cutting a bright ridge of emission regions into two parts. H$\alpha$ emission
is associated with the bright ridge of emission regions seen in the $V$ and $I$
images, and also with another emission patch across the dust lane coming in
from the northwest. The emission ridge has $V-I$ colors around 1.0, but in the
surrounding dusty area the $V-I$ colors are as red as 2. The southwestern
emission region across the dust lane from the emission ridge has a blue $V-I$
color of 0.5.  The nuclear spectrum of NGC\,6621 has been classified to be of
starburst type \citep{vlx95}. The ground-based near-infrared data of
\citet{bus90} show that the nucleus  likely lies near the ridge of bright
emission seen in  our optical images, but it is not possible to constrain the
true nuclear location from these images.

\begin{figure*}[th]
\centering
\includegraphics[width=6in,angle=270]{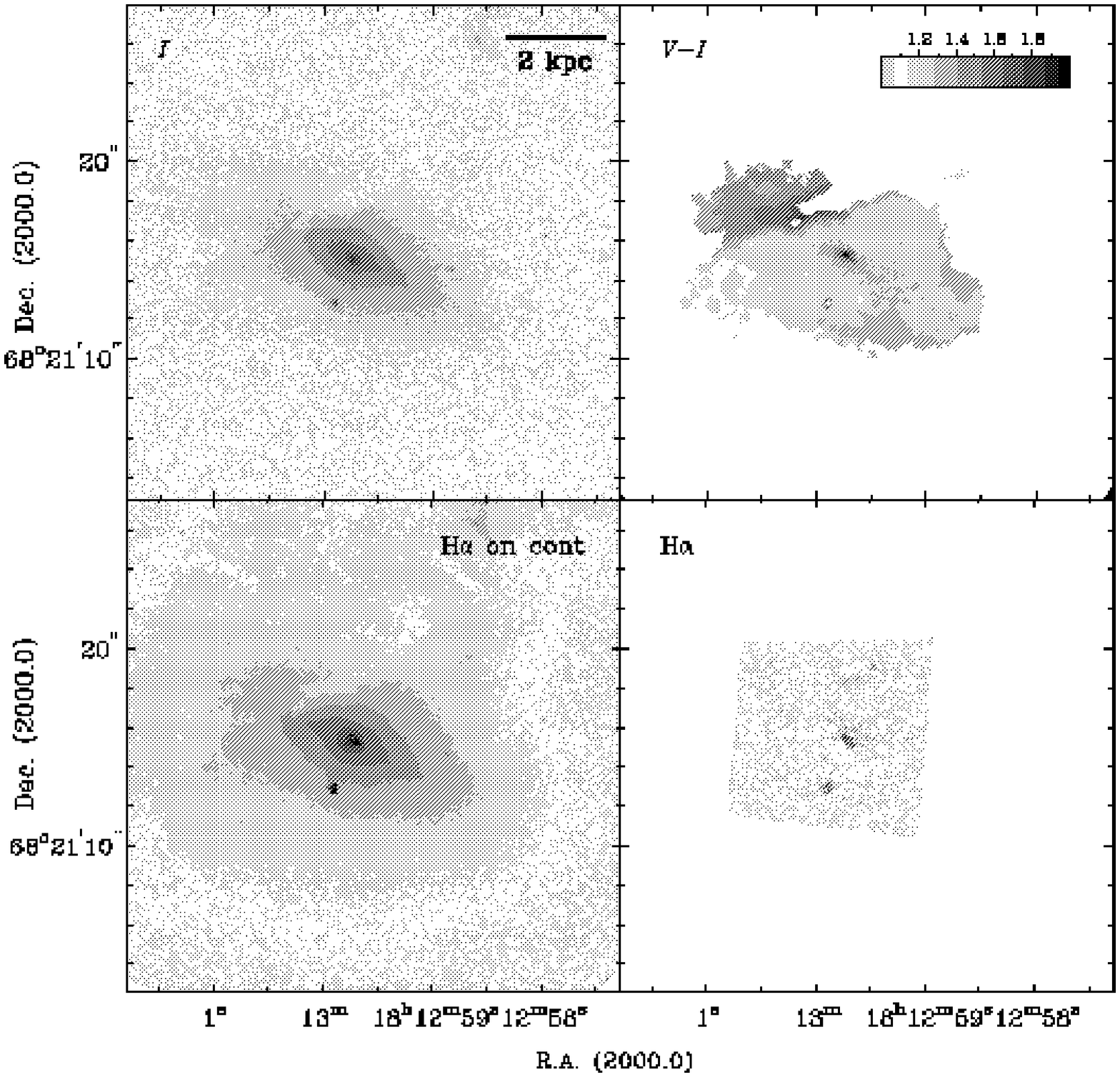}
\caption{NGC\,6622, as in Figure~\ref{f:n4038}. The H$\alpha$ contour level in the 
lower left image is at 50\% of 7.3$\times$10$^{-17}$ ergs~s$^{-1}$~cm$^{-2}$ per 
pixel.\label{f:n6622}}
\end{figure*}

NGC\,6622 has a well-defined nucleus and either a strong stellar bar, or a 
close to edge-on orientation to the line-of-sight (Fig.~\ref{f:n6622}). The
$J$-band infrared  image of \citet{bus90} shows that the outer isophotes are
not too far from  being circular, making it unlikely that this is an edge-on
galaxy. There is  an obvious dust lane on the northern and northeastern side of
the galaxy,  most likely signifying perturbations caused by NGC\,6621. There
is  practically no H$\alpha$ emission associated with NGC\,6622, suggesting
that  it is an early Hubble-type galaxy. The nucleus and the bar have red
$V-I$  colors up to 2.3. The typical color in the remainder of the disk is
1.3.  There is a star forming region in the area between NGC\,6621 and
NGC\,6622  which has a blue $V-I$ color with typical values around 0.8 (near
the upper right edge of Figure~\ref{f:n6622}).

\subsection{NGC\,3509}
\label{ss:ngc3509}

\begin{figure*}[th]
\centering
\includegraphics[width=6in,angle=270]{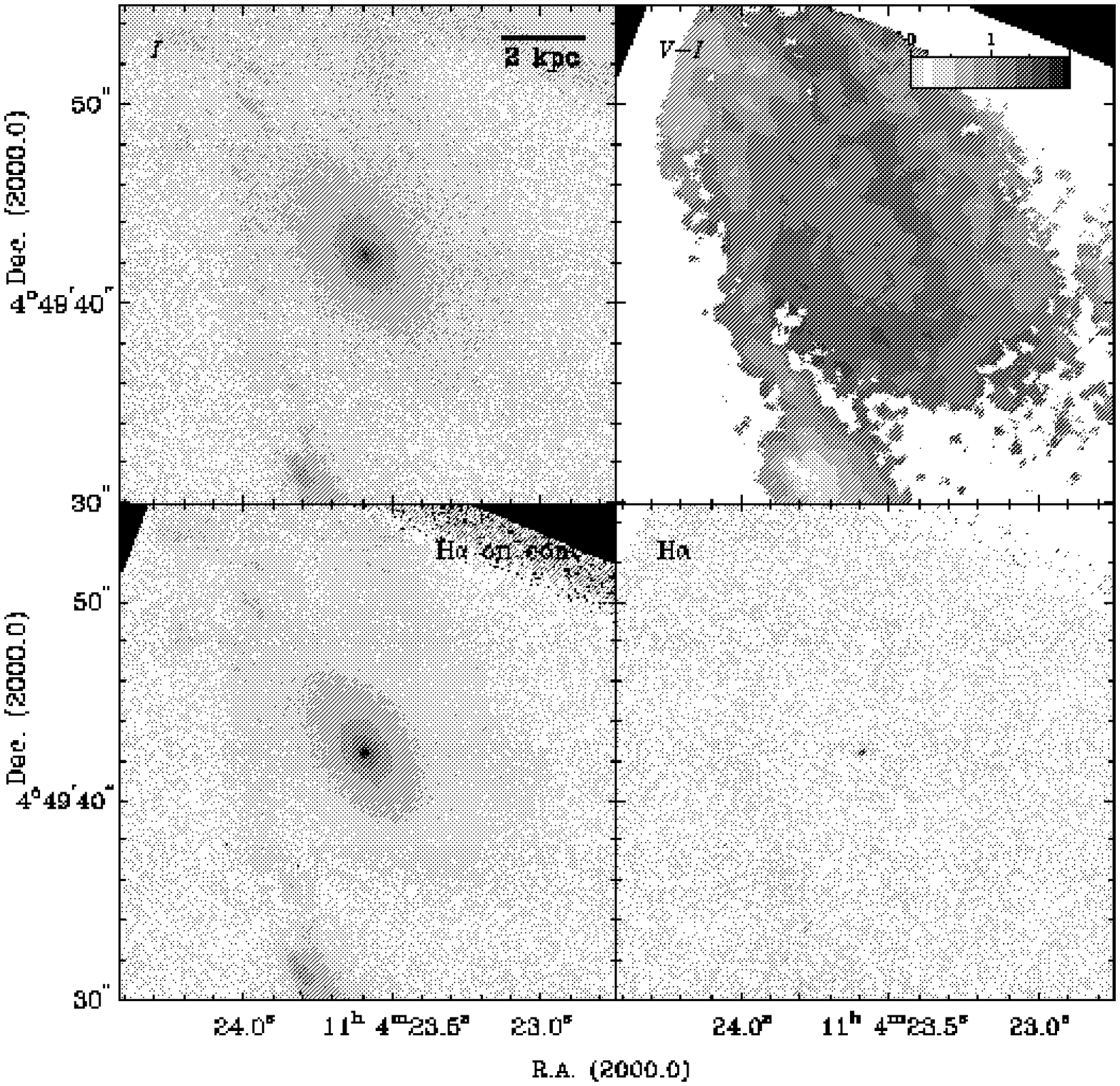}
\caption{NGC\,3509, as in Figure~\ref{f:n4038}. The H$\alpha$ contour level in the 
lower left image is at 20\% of 2.5$\times$10$^{-16}$ ergs~s$^{-1}$~cm$^{-2}$ per 
pixel.\label{f:n3509}}
\end{figure*}

The second-most distant galaxy of this sample, NGC\,3509, falls in the middle
of the Toomre Sequence. The sketch of this system by \citet{tmr77} suggests
that he envisioned a large tail curving to the northwest and a shorter tail
extending to the southwest. Deep ground-based CCD imagery obtained by one of us
(JEH) suggests that the southeastern feature is not a tail, but rather the
bright ridge of an inclined disk. There is a very bright rectangular-shaped
feature at the head of the southeastern feature, directly south of a central
bulge \citep*[see also][]{arp66}, which may have been interpreted as the bulge
of the second system.  Our HST/WFPC2 broad-band imagery shows that this region
resolves into a number of bright, blue stellar associations
(Figure~\ref{f:n3509}), and we believe it is much more likely to be a
collection of bright star forming regions within the perturbed disk than the
nucleus of a second system. The extremely blue colors of this region ($V-I$
colors around 0.5; the tip of this region is seen at the bottom center of
Figure~\ref{f:n3509}) are in spectacular contrast to the rest of the galaxy
(typical $V-I$ values from 1.1 to 1.4), and indicate active star formation.

In our images we can clearly identify a single relatively undisturbed
nucleus, surrounded by a swirl of dust lanes (center of
Figure~\ref{f:n3509}). The H$\alpha$ image reveals a peculiar
double-peaked H$\alpha$ structure straddling the nucleus in an
orientation which is perpendicular to the main body of the nuclear
region. H$\alpha$ emission, indicating ongoing star formation, is
also seen in the blue region south of the nucleus.

In the absence of an obvious second tidal tail in the ground-based imagery, we
find little evidence that NGC 3509 is the obvious result of a major disk-disk
merger. It may instead be the result of a minor merger, or an ongoing
interaction with a smaller companion. In support of the minor merger
hypothesis, we note the appearance of a roughly circular diffuse red
concentration just to the north of the nucleus (12\arcsec~$\sim$ 6~kpc; see
Fig.~\ref{f:mosaic}, just off of the PC chip) which could be the remains of a
smaller galaxy. In support of the interaction hypothesis, on a deep R-band CCD
image taken by one of us (JEH) we note the presence of a compact object lying
$\sim$~2\arcmin\ towards southeast, along the minor axis of this system. This
possible companion also has faint low surface brightness features pointing both
toward and away from NGC 3509. 

\subsection{NGC\,520}
\label{ss:ngc520}

\begin{figure*}[th]
\centering
\includegraphics[width=3in,angle=270]{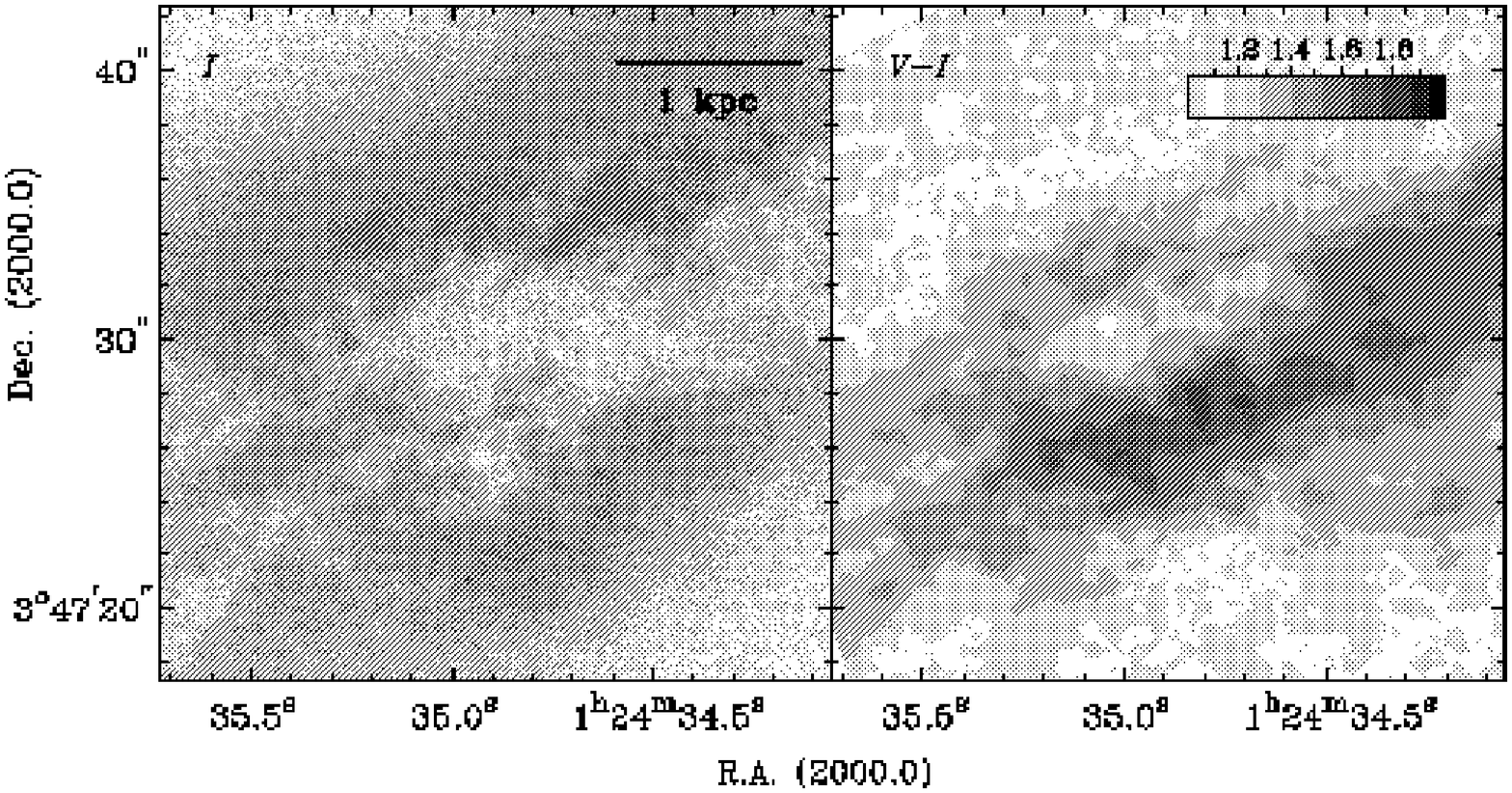}
\caption{NGC\,520 Nuc 1. Left: $I$ image. Right: $V-I$ image. The H$\alpha$ emission 
around the nucleus is not shown as it is not significantly detected above the noise 
of the image.\label{f:n520n1}}
\end{figure*}

The southeastern component of NGC\,520 is hidden behind a prominent and
intricate  dust lane (Figure~\ref{f:n520n1}). This dust is likely associated
with the dense edge-on molecular disk imaged in CO (1--0) by
\citet{san88b} and \citet{yun01}. Adopting standard conversion factors, the  observed peak
CO flux density suggests that the nucleus of NGC 520 is  hidden beneath
$A_V\sim~$300 magnitudes of extinction. It is thus not surprising that the 
position of the primary nucleus is impossible to determine from our data.
High-resolution future NICMOS observations  may point out where this nucleus
lies, but based on earlier ground-based near-infrared images
\citep[e.g.,][]{bus92,kot01} the nucleus likely lies in the center of the 
heaviest dust absorption. Radio continuum images \citep*[e.g.,][]{car90} show a
disk-like morphology coincident with the CO disk, suggesting on-going star
formation within the central molecular disk.  Practically no H$\alpha$ emission
is seen in our H$\alpha$ image at the position of the primary nucleus. Red
$V-I$ colors are seen along most of the primary nucleus. However, at the
location of the most dust-affected regions, as seen in the $I$ image, we see
relatively blue colors of 0.8 in the color index image. Again, as in the case
of NGC\,4676A, we interpret this as scattering of emission from young stars,
hinted at in the $V$-band image (not shown here).

\begin{figure*}[th]
\centering
\includegraphics[width=6in,angle=270]{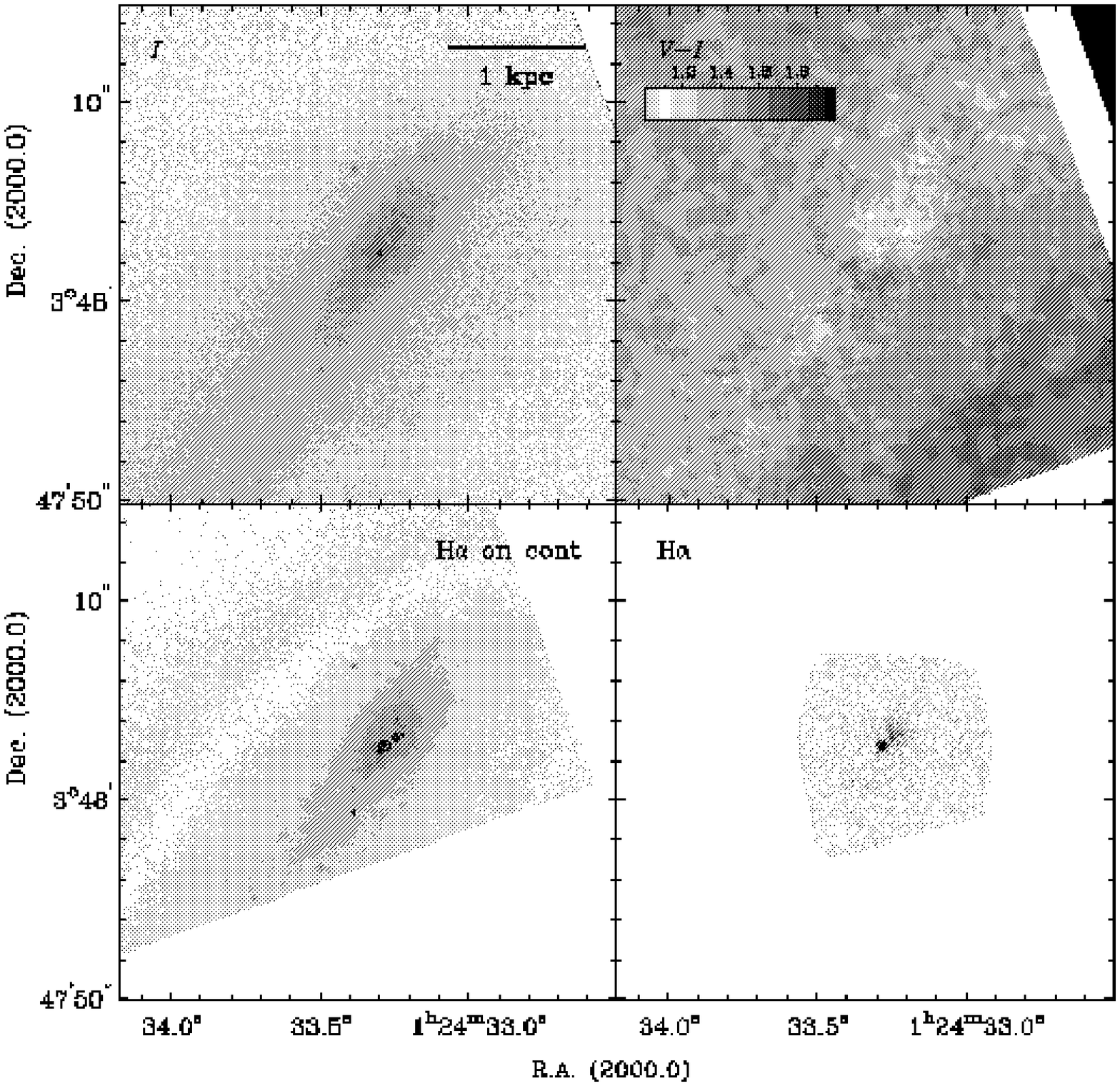}
\caption{NGC\,520 Nuc 2, as in Figure~\ref{f:n4038}. The H$\alpha$ contour levels 
in the lower left image are at (20\%, 90\%) of 1.4$\times$10$^{-16}$ 
ergs~s$^{-1}$~cm$^{-2}$ per pixel.\label{f:n520n2}}
\end{figure*}

The secondary, northwestern nucleus (Fig.~\ref{f:n520n2}) comes presumably from
an earlier Hubble-type disk galaxy that is now embedded within an extended
atomic gas disk associated with the primary galaxy  \citep{hib96}. The
secondary nucleus is very well defined, although  its surrounding disk seems to
have been mostly disrupted in the merger, and  probably contributes now to the
extended optical tails.  The only significant H$\alpha$ emission near the
secondary nucleus is in clumps within 1\arcsec~(150 pc), and displaced to the
northwest. The secondary nucleus appears to have a bluish color, with $V-I$ of
0.6. This nucleus is observed to be in a  post-starburst  phase
\citep{stan90,stan91b,ber93}. The $V-I$ color towards the northwest of  the
nucleus has bluish values around 0.8, but dust reddens the color to 1.2 at 
about 3\arcsec~(450 pc) southeast of the nucleus.

\subsection{NGC\,2623}
\label{ss:ngc2623}

\begin{figure*}[th]
\centering
\includegraphics[width=6in,angle=270]{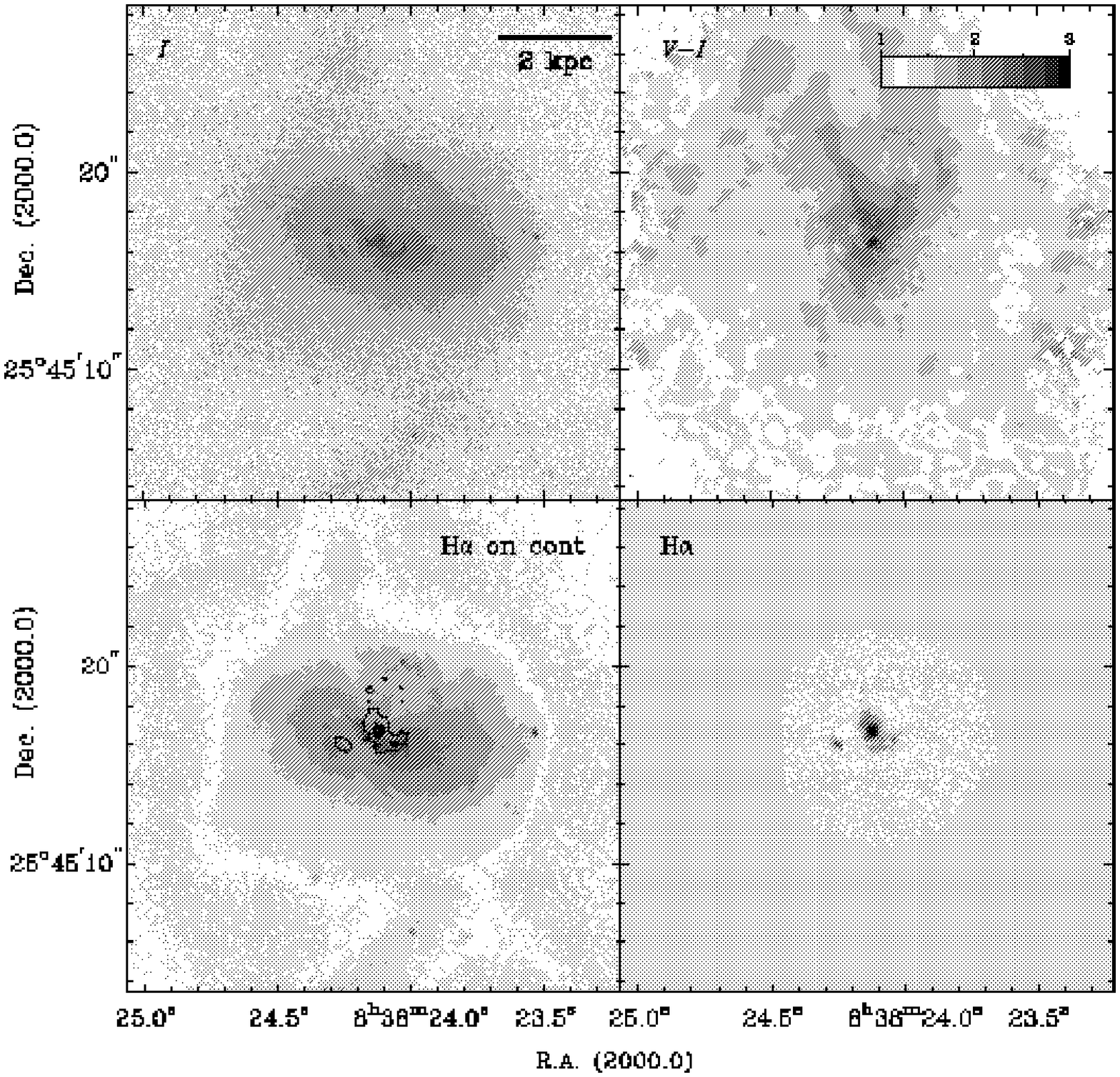}
\caption{NGC\,2623, as in Figure~\ref{f:n4038}. 
The H$\alpha$ contour levels in the lower left image are at (10\%, 50\%, 80\%, 90\%) 
of 3.3$\times$10$^{-16}$ ergs~s$^{-1}$~cm$^{-2}$ per pixel.\label{f:n2623}}
\end{figure*}

Consistent with its classification as a fairly advanced merger system, only one
galaxy body is evident in NGC\,2623 (Fig.~\ref{f:n2623}). Again,  based on the
high spatial resolution optical {\it HST} images it is impossible to tell the
exact location of the nucleus or nuclei. A comparison to the $JHK$ NICMOS
images published by \citet{sco00} suggests that the position of the brightest
optical peak in the center of the main body of NGC\,2623 is the true position
of the nucleus, near J2000.0 coordinates of R.A.=8$^{\rm h}$38$^{\rm m}$24\fs
1~and Dec.=25$\degr$45$\arcmin$16\farcs 6. This nucleus is classified as a
LINER by \citet{daha85} and \citet*{vacci98}. Dust surrounds the nucleus mostly
on the southeastern, northern, and northwestern sides, although numerous dust
lanes criss-cross the whole nuclear region (see Fig.~\ref{f:n2623}).

A study by \citet{joy87} confirms the coalescence of two galaxies. Based on
their near-infrared observations, they conclude that the merger is complete,
because only one nucleus is visible (which is surrounded by a single, symmetric
nuclear region).  This impression is confirmed by CO observations, which reveal
a single centrally concentrated compact molecular disk with simple rotational
kinematics \citep{bry99}. There are also numerous young star clusters scattered
around the main body and the surrounding shreds.  

Most of the H$\alpha$ emission comes from a disk around the nuclear position 
and from a location north of the nucleus, but also from a position in the dust
patch southeast of the nucleus. The nucleus has a very red $V-I$ color at 2.5,
consistent with the reddish near-infrared colors found for the nucleus by
\citet{sco00}. The red near-infrared $H-K$ color, together with a relatively
large CO  absorption index value \citep{ridg94} suggest that there has been a
recent starburst in the nuclear region.  The red colors continue for about
2\arcsec~(720 pc) to the north and  northwest. The main body of NGC\,2623 has
$V-I$ colors close to 1.0 or just  below it.

\subsection{NGC\,3256}
\label{ss:ngc3256}

\begin{figure*}[th]
\centering
\includegraphics[width=6in,angle=270]{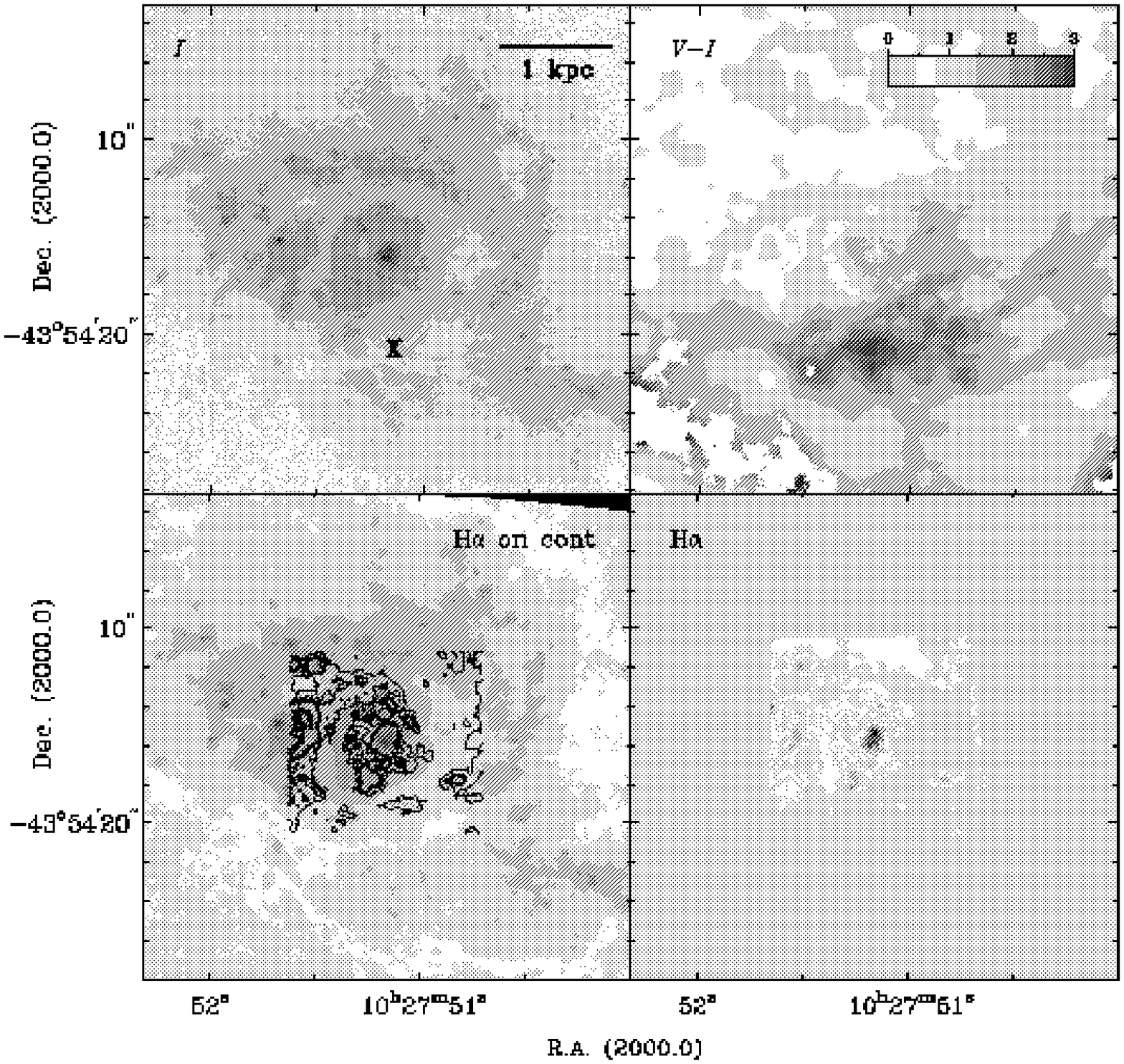}
\caption{NGC\,3256, as in Figure~\ref{f:n4038}. The location of the putative second 
nucleus is marked with an ``X'' in the upper left panel. The H$\alpha$ contour 
levels in the lower left image are at (10\%, 50\%, 80\%, 90\%, 95\%, 98\%) of 
3.5$\times$10$^{-16}$ ergs~s$^{-1}$~cm$^{-2}$ per pixel.\label{f:n3256}}
\end{figure*}

Our optical {\it HST} images show one obvious nucleus located in the center of
Figure~\ref{f:n3256} (see also \citealt{zepf99}). This nucleus is surrounded by
a relatively symmetric spiral morphology. Near-infrared, radio continuum, and
X-ray  observations \citep[e.g.,][]{nor95,kot96,lira02} have discovered a
second nucleus about 5\arcsec~(885 pc) south of the primary nucleus (marked
with an ``X'' in Fig.~\ref{f:n3256}), presumably hidden beneath the dark (red)
dust feature directly south of the primary nucleus in the $V-I$ image of
Fig.~\ref{f:n3256} (see also \citealt{eng03}). We do not see an optical source
at the position of the second near-infrared, radio, and X-ray source. Our data
therefore do not shed any new light on the nature of this source. Neither
nucleus shows any evidence of being active. Studying the question of how many
nuclei exist in this system (even three nuclei have been suggested by
\citealt{lip00}), would benefit from a detailed dynamical simulation of this
merger system. The measurements for NGC\,3256 in this paper refer only to the
one nucleus we see in our WFPC2 images.

Archival NICMOS near-infrared images suggest that the position of the primary
nucleus is close to the location of the brightest emission in our F814W image
\citep{lip00}. However, the optical {\it HST} images alone only show that  this
position is part of a ring-like structure, and there is nothing noticeably
different about this location other than that it is the brightest clump in the 
image. Other bright regions are likely to be young star clusters or globular 
clusters, as reported by \citet{zepf99}. 

The H$\alpha$ emission is extended and follows the spiral structure, but the
brightest emission occurs at the position of the primary nucleus. Most 
previous studies regard this nucleus to be of starburst type, although 
\citet{ver86} were unable to resolve its true nature. While the primary
nucleus has a blue $V-I$ color near 0.7, the dusty region covering the 
secondary nucleus to the south of the primary nucleus  has a $V-I$ color of 2.5
or redder. In contrast, the spiral arms visible in the $V$ and $I$ images have
a blue $V-I$ color between 0.5 and 1.0.

\subsection{NGC\,3921}
\label{ss:ngc3921}

\begin{figure*}[th]
\centering
\includegraphics[width=6in,angle=270]{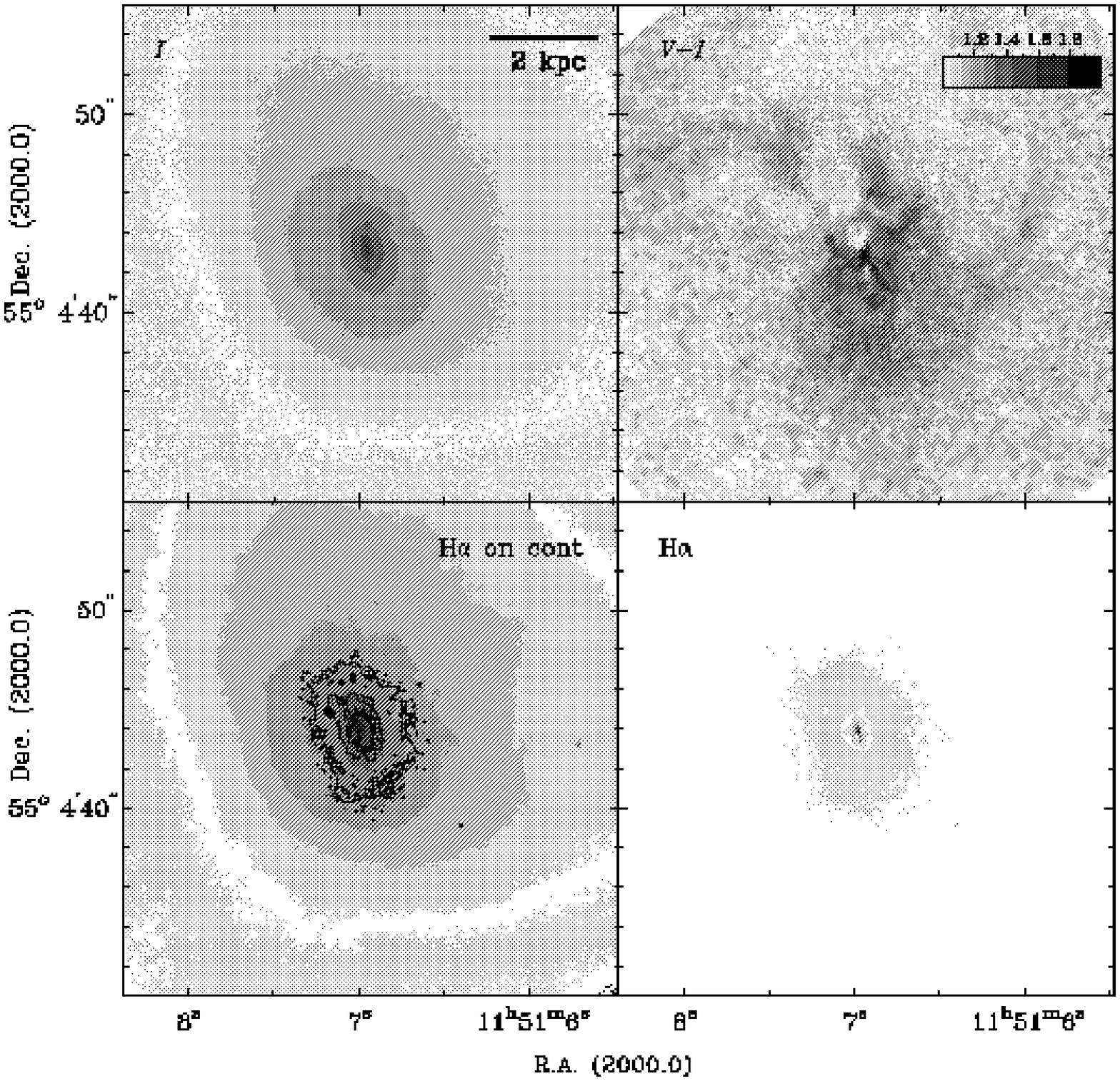}
\caption{NGC\,3921, as in Figure~\ref{f:n4038}. The H$\alpha$ contour levels in the 
lower left image are at (10\%, 30\%, 50\%, 70\%, 90\%) of 1.9$\times$10$^{-16}$ 
ergs~s$^{-1}$~cm$^{-2}$ per pixel.\label{f:n3921}}
\end{figure*}

Only one nucleus is visible in NGC\,3921 (Figure~\ref{f:n3921}). This galaxy is
a post-merger system that has developed an elliptical-like central region
\citep[see, e.g.,][]{schw96a,schw96,hib99}, although numerous dust lanes still 
criss-cross this area (see Fig.~\ref{f:n3921}). The evidence for a recent
merger  can be seen in the outer region where tails and  plumes prevail
\citep{hib96}. Several compact open star clusters or globular clusters can also
be seen in  the {\it HST} optical images kindly provided by B.~Whitmore 
\citep[see also][]{schw96}.

The  H$\alpha$ emission is centered in a disk around the nucleus, but extends
to larger radii. The nucleus is classified as a LINER by \citet{sta82a,sta82b}
and  \citet{daha85}. It lies in a location where the $V-I$ color gradient is 
very steep as seen in the $V-I$ image, but the nucleus itself has
normal values of $V-I$ around 1.1. To the southwest of the nucleus the $V-I$
color increases to $\sim$~1.8 at about 0\farcs 5~(190 pc) distance from the
nucleus, and then decreases and stays around 1.1--1.2 on that side of the
galaxy. To the northeast, the $V-I$ color decreases to values as low as 0.7 at
about 0\farcs 5 from the nucleus, then stays around 0.9 for another 1\farcs
5~(570 pc). Dust extinction is at least partly responsible for this effect.
\citet{schw96a} noticed that the nucleus lies substantially off-centered from
the outer stellar isophotes of the main body.

\subsection{NGC\,7252 (``Atoms--for--Peace'' Galaxy)}
\label{ss:ngc7252}

\begin{figure*}[th]
\centering
\includegraphics[width=6in,angle=270]{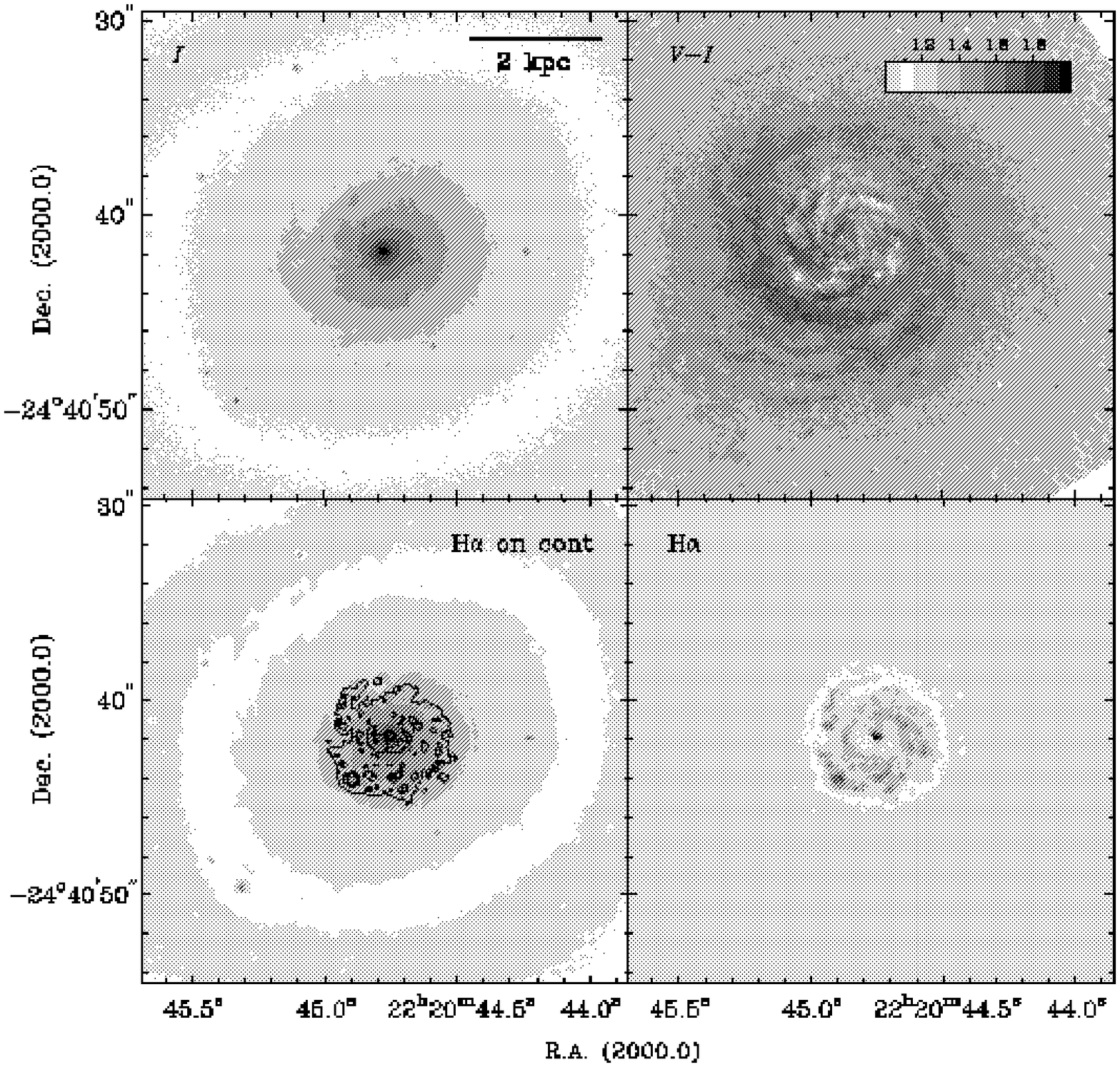}
\caption{NGC\,7252, as in Figure~\ref{f:n4038}. The H$\alpha$ contour levels in the 
lower left image are at (10\%, 30\%, 50\%) of 7.2$\times$10$^{-16}$ 
ergs~s$^{-1}$~cm$^{-2}$ per pixel.\label{f:n7252}}
\end{figure*}

We use the  {\it HST} $V$ and $I$ images from \citet{mil97} to study the most
advanced merger in the Toomre Sequence.  The remnant from the merger is
probably between 0.5 and 1 Gyrs old  \citep*{sch82,hib95b}. It has a single
nucleus and is characterized by a $r^{1/4}$ profile typical of elliptical
galaxies \citep*{sch82,mil97,hib99}. It was quite a surprise, therefore, when
{\it HST} revealed a central spiral disk (\citealt{mil97} and
Fig.~\ref{f:n7252}). The disk extends to a radius of about 6\arcsec~(1.8\,kpc),
and is therefore coincident with the molecular disk imaged by \citet{wang92}.
Our new {\it HST} H$\alpha$ image of this galaxy shows the disk to be actively
forming stars, while the $V-I$ color map shows dust to be intertwined between
the arms. In light of the results from recent hydrodynamical simulations 
\citep{bar02}, and the observation of tidal gas streaming back into the central
regions of this remnant from the tidal tails \citep{hib94, hib95b}, it seems
likely that this central disk structure results  from the re-accretion of
tidally raised gas into the now-relaxed central potential. 

The nucleus appears resolved, and is very bright, as we show in
Section~\ref{s:mergerstage}. H$\alpha$ emission largely follows the spiral
arms, although there appears to be a central H$\alpha$ source as well. 
However, our continuum subtraction, unfortunately, is uncertain in this area
due to an optically very bright nucleus. Similarly, the central $V-I$ color is
difficult to determine reliably due to the same problem, but the color there
appears to be close to 1.3, whereas the surrounding spirals have $V-I$ colors
as blue as 0.5.

Recent X-ray observations with ASCA revealed X-ray emission that can be
described by a two-component model \citep{awa02}, with soft emission (kT = 0.72
keV), and a hard component (kT $>$ 5.1 keV).  The hard component might indicate
the existence of nuclear activity or even an intermediate mass black hole.

\section{CHANGE IN PHOTOMETRIC PROPERTIES WITH MERGER STAGE}
\label{s:mergerstage}

We performed aperture photometry of the Toomre Sequence galaxies to better
understand the evolution of the nuclear region of merging galaxies. Because the
isophotes in the nuclear region of merging galaxies are highly twisted, we did
not attempt to fit ellipses and obtain radial surface brightness profiles.
Instead we performed aperture photometry within circular apertures, with radii
of 100 pc and 1 kpc. The lower value was dictated by the smallest resolvable
physical scale in the most distant of our sample galaxies, whereas the upper
value is somewhat arbitrary, but set because of our emphasis on the properties
of the nuclear region. We have already characterized the morphology of the
nuclear region at {\it HST} resolution in the individual galaxies above. Here
we only report on the bulk photometric nuclear properties of each identified
merger component galaxy.

Visual inspection of the images quickly reveals that none of the galaxies in
the sample possess an extremely bright, point-like active nucleus. Based on
spectroscopic evidence, as cited in Section~\ref{s:images}, the only mildly
active nuclei are in NGC\,7592A (Seyfert 2), NGC\,4676 Nuc~1 and Nuc~2
(LINER), NGC\,2623 (LINER), and NGC\,3921 (LINER). It is difficult to assess
what the exact contribution of the nucleus to the integrated luminosities is.

The photometric measurements, converted to luminosity densities, are presented
in Figures~\ref{f:iphot}--\ref{f:haphot}. The ``merger stage'', plotted along
the x-axis, refers to the  original order of the sequence given by
\citet{tmr77}, where 1 corresponds to NGC\,4038/39 (the earliest merger stage)
and 11 to NGC\,7252 (the latest merger stage). The exact correspondence is
given in the figure caption to Figure~~\ref{f:iphot}.   Our observations do not
suggest an obvious reordering of this sequence, although we are hampered by our
inability to unambiguously identify all of the putative nuclei. Our tentative
impression is that NGC\,7764A (merger stage=4) may be interacting with the
disturbed galaxy lying off of our image, in which case it belongs near the
beginning of the sequence. And NGC\,3509 (merger stage=6) is not an obvious
merger at all. However, we will await NICMOS observations of this sequence 
before reaching any firm conclusions on this matter. We note that none of our
conclusions will be affected if these two systems move earlier along or off of
the sequence. 

Apart from the large scatter in Figure~\ref{f:iphot} we see that the three
latest merger stage systems (NGC\,3256, NGC\,3921, and NGC\,7252) have the
highest intrinsic luminosity densities. This is true for both apertures that we
used (100 pc and 1 kpc). The range of luminosity density within 100 pc
corresponds to a factor of more than 300. To a large extent this reflects the
varying dust extinction in the central region, but partly also the morphology
of the nuclear region (well-defined nuclei are bright, whereas components with
no clearly recognizable nucleus are faint). With only $V-I$ colors available it
is difficult to make a meaningful estimate of dust extinction. But, for
example, Figure~\ref{f:viphot} shows that in NGC~2623 the integrated $V-I$
color is very red within a 100 pc radius aperture, with a value around 2.2.
Assuming that the true $V-I$ color is 1.4 (a more typical value seen in the
same figure), would imply almost $A_{V}$ = 2 mag of extinction with a typical
extinction law of A$_{\lambda}\sim \lambda^{-1.85}$ \citep{lan84}. Since the
extinction in the $K$-band is only one tenth of that in the $V$-band, our
future NICMOS images offer some hope of revealing the true location of the 
nucleus.

While the $V-I$ colors for the 100~pc aperture attain values anywhere between
0.8 and 2.4, the majority of the nuclei have rather typical colors for the
centers of spiral nuclei and elliptical galaxies, ranging from 1.2 to 1.5
\citep[cf., e.g.,][]{car97}. The presumably  latest stage merger, NGC\,7252,
has colors even bluer than normal ellipticals. However, the morphology of the
dust lanes we see in the images suggests that dust obscuration is likely to be
significant in many systems of the Toomre Sequence.

The H$\alpha$ + [\ion{N}{2}] luminosities (Figure~\ref{f:haphot}) show a
marginal trend of increasing luminosities towards the late-stage mergers, most
obviously in the largest 1~kpc aperture. However, the scatter is large, and
based on the sample of 15 nuclei it is impossible to establish any definite
trends.

Since mergers are expected to bring material into the center of galaxies and
generate star formation there, we also study the nuclear concentration of light
and line emission along the merging  sequence. We measure the nuclear
concentration quantitatively by taking the ratio of the luminosities within 100
pc to that within 1 kpc, for both the $I$-band (Figure~\ref{f:irat}) and the
line emission (Figure~\ref{f:harat}). We discuss the meaning of these ratios
briefly in the following section.

\section{DISCUSSION AND SUMMARY}
\label{s:discussion and summary}

The nuclear morphologies and central concentration are expected to be strongly
affected by the timing and intensity of induced inflows and star formation
during the merger. Numerical models of interacting galaxies have illustrated
one mechanism by which interactions trigger these inflows and nuclear activity.
Shortly after the galaxies first collide, the self-gravity of each disk
amplifies the collisional perturbation into a growing $m=2$ mode
\citep{nog88,bar91,hos96}. Depending on the structural properties of each
galaxy, this mode can  take the form of a strong bar, prominent spiral arms,
or  a pronounced oval distortion  \citep{hos96,hos97}. These non-axisymmetric
structures drive shocks in the interstellar medium of the galaxies, leading to
a spatial offset between the gaseous and stellar components which robs the gas
of angular momentum, and can drive it inwards well before the galaxies
ultimately merge \citep{bar91,her95,hos96}.  The onset, strength, and duration
of starbursts in interacting pairs can vary widely, depending on the properties
of the host galaxies. Ultimately, however, any dynamical stability provided by
the presence of a central bulge or diminished disk surface density will be
overwhelmed by the strong gravitational torques and dissipative shocks during
the final merging of the pair, at which point gaseous inflow and central
activity should be ubiquitous. Once the coalescence of the galaxies is
complete, the gravitational potential settles down and gas can resettle into
nuclear and/or spatially extended disks, supporting an extended period of
relatively quiescent star formation.

\begin{figure*}
\centering
\includegraphics[width=2.5in,angle=270]{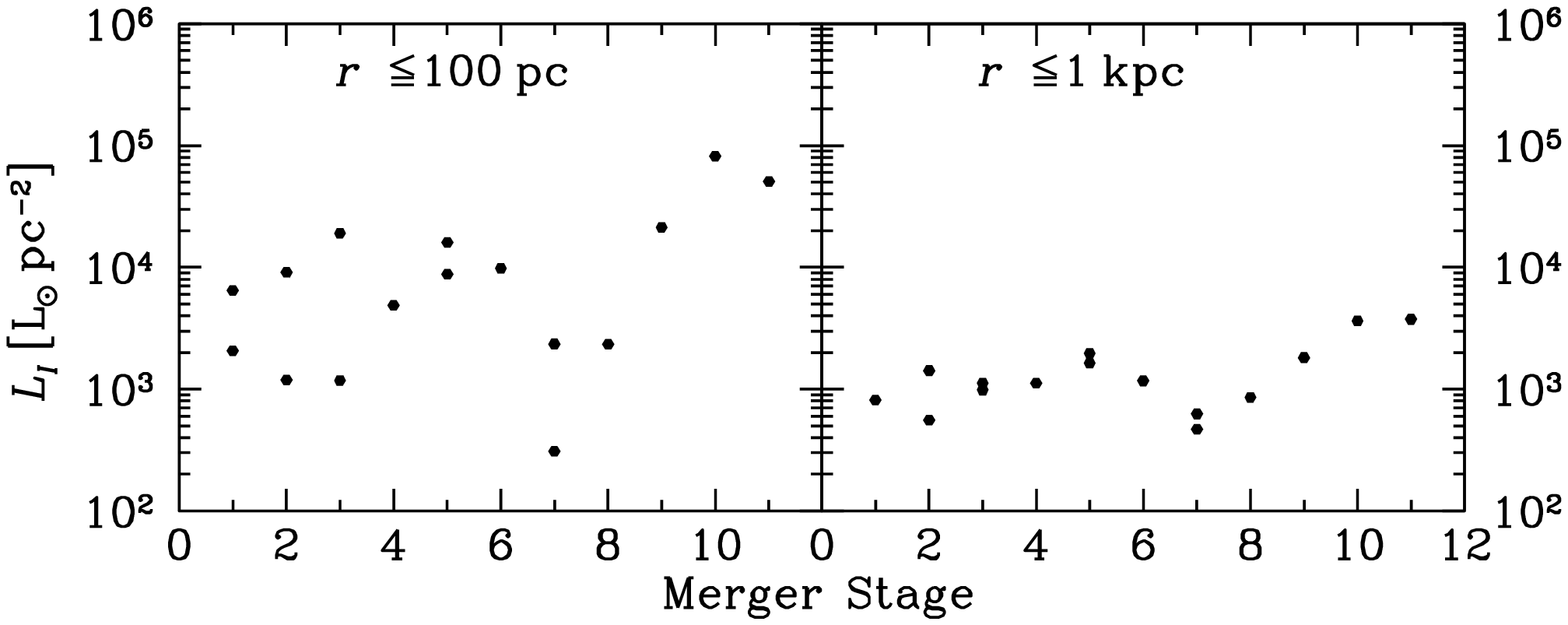}
\figcaption{Circular $I$-band aperture photometry centered on the putative nucleus 
in each of the merging galaxies of the Toomre Sequence. The horizontal axis 
represents the sequence as originally conceived by \citet{tmr77}. The merging stage 
numbers correspond to the individual galaxies as follows: 1: NGC\,4038/39; 2: 
NGC\,4676; 3: NGC\,7592; 4: NGC\,7764A; 5: NGC\,6621/22; 6: NGC\,3509; 7: NGC\,520; 
8: NGC\,2623; 9: NGC\,3256; 10: NGC\,3921; 11: NGC\,7252. The two-dimensional 
projected luminosity density is presented on the vertical axis, within apertures 
specified within the frame. Left, within a 100  pc radius aperture, right, within 
a 1~kpc radius aperture. The luminosity densities have not been corrected for 
extinction. Uncertainties in the plotted values are smaller than the symbol 
size.\label{f:iphot}}
\end{figure*}

Based on these kinds of models, we might expect to see an evolutionary trend
for the nuclear properties of the Toomre Sequence, from quiescent star
formation in the early stages of the sequence to strong nuclear activity in the
late stages. However, we see little evidence for such trends in the morphology,
luminosity density, or colors of the stellar component, or in the morphology or
intensity of the  ionized gas emission, apart possibly from the fact that the
latest-stage systems have some of the highest broad-band luminosity densities
and hints of starbursting central gaseous disk structures. This last suggestion
is further strengthened by the trend towards bluer colors among  the three
latest-stage merger systems in Figure~\ref{f:viphot}. A trend towards bluer
colors was also seen in a study of merger remnants by \citet{marel00}.
Furthermore, the latest-stage merger systems have some of the highest nuclear
H$\alpha$ luminosities (Figure~\ref{f:haphot}) and possess some of the largest
concentrations of light in the central region (Figure~\ref{f:irat}). Such a
concentration is also seen to a certain extent in the H$\alpha$+[\ion{N}{2}]
emission (Figure~\ref{f:harat}).

\includegraphics[width=3.0in]{Laine.fig18.ps}
\figcaption{Integrated $V-I$ color indices within the circular
apertures cited within the frames, as a function of the merger stage. The
merger stage is defined as before in Figure~\ref{f:iphot}. The error bars are
based on the combined errors in the $V$- and $I$-band aperture magnitudes,
estimated using the standard propagation of error formulas.\label{f:viphot}}

\vskip 0.1in

Are our observations of the Toomre nuclei consistent with the broad picture of
interaction-induced nuclear activity in galaxies evidenced in numerical
simulations? In a broad sense, yes, although this consistency is mainly due to
the variety of dynamical responses available to interacting systems. The late
stage mergers seem to have settled down, showing evidence for nuclear
emission-line disks (NGC\,2623, NGC\,3256, NGC\,3921, and NGC\,7252),  more
concentrated luminosity profiles and a trend towards bluer colors. Toomre
Sequence objects at earlier stages show very diverse properties, with no clear
trends along the sequence. However, selecting a merging sequence based largely
on the large-scale morphology of the tidal tails biases the sample towards a
specific type of interaction  -- prograde interactions between disk galaxies --
but not a specific type of  disk galaxy.  The galaxies which make up the Toomre
Sequence likely possess a variety of structural properties so that there is no
one-to-one correspondence between observed merger stage and nuclear morphology
and activity. Put differently, the Toomre Sequence is {\it not} a true
evolutionary sequence on anything but the largest scales. Instead, it is
composed of a variety of galaxies progressing down varied paths of
interaction-driven activity. Expecting to discern a clear trend in nuclear
properties is likely a naive hope.

\includegraphics[width=3.0in]{Laine.fig19.ps}
\figcaption{Integrated H$\alpha$+[\ion{N}{2}] luminosities in units of
ergs~sec$^{-1}$ within the circular apertures as cited within the frames, as a
function of the merger stage. The merger stage is defined as before in 
Figure~\ref{f:iphot}. It was impossible to measure the flux within 1 kpc reliably 
for NGC 520 Nuc 2 (merger stage 7), since the galaxy was so close to the edge of 
the chip. Therefore, no 1 kpc luminosity is plotted here and in 
Figure~\ref{f:harat}. Uncertainties in the fluxes are no better than 50\%.
\label{f:haphot}}
\vskip 0.1in

Aside from the variety of physical conditions sampled by the Toomre nuclei,
searching for trends along the Toomre Sequence is also hampered by a number of
observational concerns. First, the orientation of the galaxies with respect to
the line-of-sight and to the orbital plane of the merger complicates the
interpretation of the images. Second, extinction  has a large effect on the
observed morphologies, colors, and  luminosities, as shown in
Section~\ref{s:mergerstage} (e.g., NGC\,520 Nuc~1 and NGC\,4676 Nuc~1). In this
context, we hope to improve upon the interpretation of the galaxy morphologies
along the sequence through an analysis of NICMOS images of the entire Toomre
Sequence. These images will suffer from much less extinction, and should also
reveal the isophotes of the stellar population more clearly, thereby helping us
in assessing the inclination of the merger components. The optical images shown
here clearly demonstrate that dust in merging galaxies, even on small scales
resolved by the {\it HST}, is not patchy enough to enable the identification of
the true nuclei in optical bands.

\includegraphics[width=2.4in,angle=270]{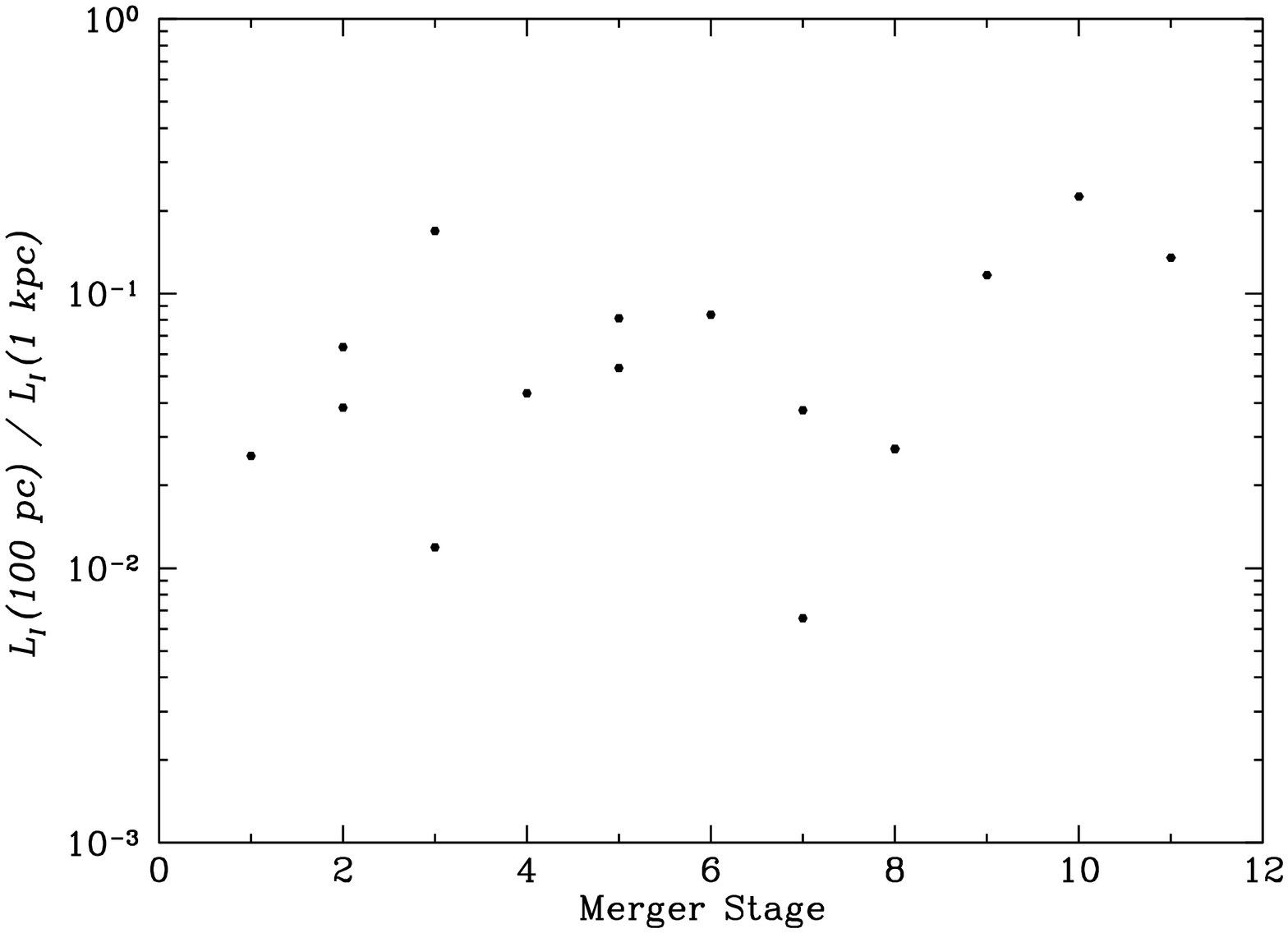}
\figcaption{Ratio of the integrated $I$-band luminosity inside 100 pc
to the integrated luminosity inside 1 kpc, plotted on a logarithmic
scale, as a function of the merger stage.\label{f:irat}}

\vskip 0.1in

In summary, these high-resolution WFPC2 images of the Toomre Sequence have
given us a detailed view of the nuclear regions of interacting and merging
galaxies. We have characterized the broad-band and emission-line morphologies
of each member of the sequence, and measured the colors and luminosity
densities of the nuclei. We find little evidence for clear trends in nuclear
properties along the merger sequence, other than a suggestive rise in the
nuclear luminosity density in the most evolved members of the sequence. The
lack of clear trends in nuclear properties is likely due both to the effects of
obscuration and geometry, as well as the physical variety of galaxies involved
in the Toomre Sequence. In subsequent papers we will combine our optical
imaging with NICMOS imaging and STIS spectroscopy of the Toomre nuclei to give
a detailed picture of the physical conditions in interacting and merging
galaxies.

\includegraphics[width=2.4in,angle=270]{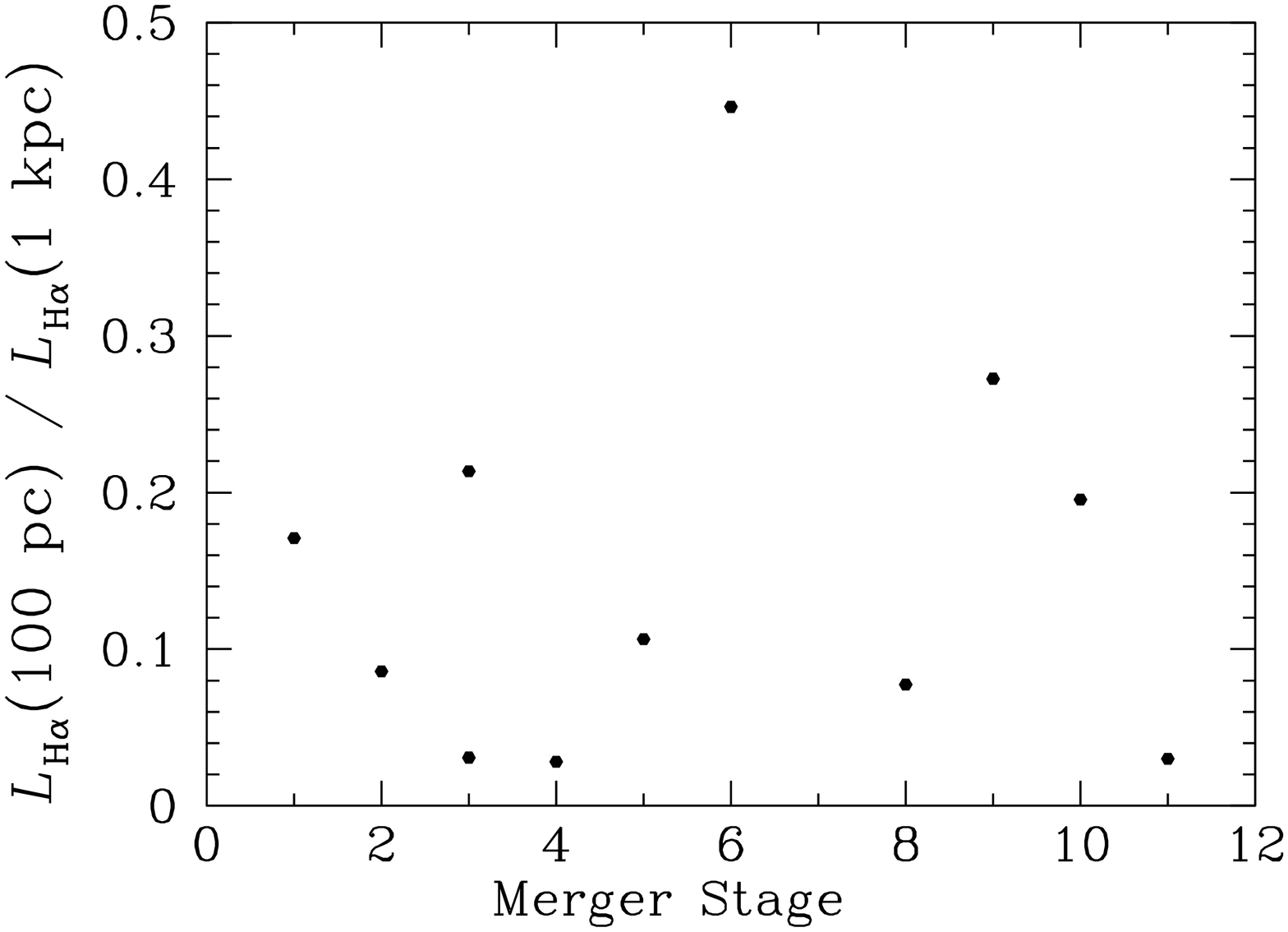}
\figcaption{Ratio of the integrated H$\alpha$+[\ion{N}{2}] luminosity inside 100 pc 
to the integrated H$\alpha$+[\ion{N}{2}] luminosity inside 1 kpc, as a function of 
the merger stage. Uncertainties in the fluxes are no better than 
50\%.\label{f:harat}}


\acknowledgments 
 
We are grateful to Dr. Brad Whitmore for providing us with fully reduced HST
WFPC2 $V$ and $I$ images of NGC\,3921, NGC\,4038/39, and NGC\,7252. We thank
Dr. Nick Scoville for providing us with his adaptive filtering code. We are
grateful to Zoltan Levay for help with the mosaiced images, and thank Dr.
William Keel for providing us with a copy of his HST image of NGC\,6621/22 for
help in planning our observations. We thank the anonymous referee for
constructive comments that improved the clarity of the paper. Support for
proposal \#8669 was provided by NASA through a grant from the Space Telescope
Science Institute, which is operated by the Association of Universities for
Research in Astronomy, Inc., under NASA contract NAS 5-26555. The research described in this paper was 
carried out, in part, by the Jet Propulsion Laboratory, California Institute of Technology, and was 
sponsored by the National Aeronautics and Space Administration.JCM acknowledges
support by the NSF through grant AST-9876143 and by a Research Corporation
Cottrell Scholarship. This research has made use of the NASA/IPAC Extragalactic
Database (NED)  which is operated by the Jet Propulsion Laboratory, California
Institute of Technology,  under contract with the National Aeronautics and
Space Administration. The LEDA database,  operated by Centre de Recherche
Astronomique de Lyon, is kindly acknowledged.


\clearpage

\begin{thebibliography}{}

\bibitem[Arp(1966)]{arp66} Arp, H. 1966, Atlas of Peculiar Galaxies,
(Pasadena: California Institute of Technology)

\bibitem[Awaki et al.(2002)]{awa02}
Awaki, H., Matsumoto, H., \& Tomida, H. 2002, \apj, 567, 892

\bibitem[Bahcall et al.(1995)Bahcall, Kirhakos \& Schneider]{bah95}
Bahcall, J. N., Kirhakos, S., \& Schneider, D. P. 1995, \apj, 450, 486

\bibitem[Barnes(1988)]{bar88} 
Barnes, J. E. 1988, \apj, 331, 699

\bibitem[Barnes(1998)]{bar98} 
Barnes, J. E. 1998, in Saas-Fee Advanced Course No.~26, 
``Galaxies: Interactions and Induced Star Formation'', ed. D.~Friedli, 
L.~Martinet \& D.~Pfenniger (Berlin: Springer--Verlag), 275

\bibitem[Barnes(2002)]{bar02} 
Barnes, J. E., 2002, \mnras, 333, 481

\bibitem[Barnes \& Hernquist(1991)]{bar91} 
Barnes, J. E., \& Hernquist, L. 1991, \apjl, 370, L65

\bibitem[Barnes \& Hernquist(1992)]{bar92} 
Barnes, J. E., \& Hernquist, L. 1992, \nat, 360, 715

\bibitem[Barnes \& Hernquist(1996)]{bar96} 
Barnes, J. E., \& Hernquist, L. 1996, \apj, 471, 115

\bibitem[Bekki \& Couch(2001)]{bekki01} 
Bekki, K., \& Couch, W. J. 2001, \apjl, 557, L19

\bibitem[Bernl\"{o}hr(1993)]{ber93} 
Bernl\"{o}hr, K. 1993, \aap, 270, 20

\bibitem[Biretta et al.(2000)]{bir20} 
Biretta, J. A., et al. 2000, WFPC2 Instrument Handbook, Version 5.0, 
(Baltimore: STScI)

\bibitem[Borne et al.(2000)]{bor00}
Borne, K. D., Bushouse, H., Lucas, R. A., \& Colina, L. 2000, \apj, 529, 77

\bibitem[Bryant \& Scoville(1999)]{bry99}
Bryant, P. M., \& Scoville, N. Z. 1999, \aj, 117, 2632

\bibitem[Bushouse \& Stanford(1992)]{bus92} 
Bushouse, H. A., \& Stanford, S. A. 1992, \apjs, 79, 213

\bibitem[Bushouse \& Werner(1990)]{bus90} 
Bushouse, H. A., \& Werner, M. W. 1990, \apj, 359, 72

\bibitem[Carollo et al.(1997)]{car97} 
Carollo, C. M., Franx, M., Illingworth, G. D., \& Forbes, D. A. 1997, \apj, 
481, 710

\bibitem[Carral et al.(1990)Carral, Turner, \& Ho]{car90} 
Carral, P., Turner, J. L., \& Ho, P. T. P. 1990, \apj, 362, 434

\bibitem[Dahari(1985)]{daha85} 
Dahari, O. 1985, \apjs, 57, 643

\bibitem[Dubinski et al.(1996)Dubinski, Mihos, \& Hernquist]{dub96} 
Dubinski, J., Mihos, J. C., \& Hernquist, L. 1996, \apj, 462, 576

\bibitem[Duc \& Brinks(2001)]{duc01} 
Duc, P. A., \& Brinks, E. 2001, in ASP Conf. Proc. 240, Gas and Galaxy 
Evolution, ed. J. E. Hibbard, M. Rupen, \& J. H. van Gorkom 
(San Francisco: ASP), 181

\bibitem[Elmegreen et al.(1993)Elmegreen, Kaufman, \& Thomasson]{elm93} 
Elmegreen, B. G., Kaufman, M., \& Thomasson, M. 1993, \apj, 412, 90

\bibitem[English et al.(2003)]{eng03}
English, J., Norris, R. P., Freeman, K. C., \& Booth, R. S. 2003, AJ, 125, 1134

\bibitem[Fabbiano et al.(1997)]{fab97}
Fabbiano, G., Schweizer, F., \& Mackie, G. 1997, \apj, 478, 542

\bibitem[Ford et al.(2002)]{for02}
Ford, H. C., Illingworth, G. D., Clampin, M., Hartig, G., the ACS Science Team 
and ESA 2002, preprint (STScI-PRC02-11d)

\bibitem[Georgakakis et al.(2000)Georgakakis, Forbes \& Norris]{geo00}
Georgakakis, A., Forbes, D.A., \& Norris, R.P. 2000, \mnras, 318, 124

\bibitem[Hattori et al.(2002)]{hat02} 
Hattori, T., Yoshida, M., Ohtani, H., Ishigaki, T., Sugai, H., Hayashi, T., 
Ozaki, S., \& Ishii, M. 2002, \pasj, 54, 393

\bibitem[Hernquist(1992)]{her92} 
Hernquist, L. 1992, \apj, 400, 460

\bibitem[Hernquist \& Mihos(1995)]{her95} 
Hernquist, L., \& Mihos, J. C. 1995, \apj, 448, 41

\bibitem[Hibbard \& Mihos(1995)]{hib95b} 
Hibbard, J. E., \& Mihos, J. C. 1995, \aj, 110, 140

\bibitem[Hibbard \& van Gorkom(1996)]{hib96} 
Hibbard, J. E., \& van Gorkom, J. H. 1996, \aj, 111, 655

\bibitem[Hibbard et al.(1994)]{hib94} 
Hibbard, J. E., Guhathakurta, P., van Gorkom, J. H, \& Schweizer, F. 1994, 
\aj, 107, 67

\bibitem[Hibbard et al.(2001)]{hib01} 
Hibbard, J. E., van der Hulst, J. M., Barnes, J. E., \& Rich, R. M. 2001, 
\aj, 122, 2969

\bibitem[Hibbard \& Yun(1999)]{hib99} 
Hibbard, J. E., \& Yun, M. S. 1999, \apj, 522, L93

\bibitem[Holtzman et al.(1995)]{hol95} 
Holtzman, J. A., Burrows, C. J., Casertano, S., Hester, J. J., Trauger, J. T., 
Watson, A. M., \& Worthey, G. 1995, \pasp, 107, 1065

\bibitem[Hutchings \& Neff(1992)]{hn92}
Hutchings, J. B., \& Neff, S. G. 1992, \aj, 104, 11

\bibitem[Joseph \& Wright(1985)]{jos85} 
Joseph, R. D., \& Wright, G. S. 1985, \mnras, 214, 87

\bibitem[Joy \& Harvey(1987)]{joy87}
Joy, M., \& Harvey, P.M. 1987, \apj, 315, 480

\bibitem[Keel et al.(1985)]{keel85} 
Keel, W. C., Kennicutt, R. C., Jr., Hummel, E., \& van der Hulst, J. M. 1985, 
\aj, 90, 708

\bibitem[Kewley et al.(2001)]{kew01} 
Kewley, L. J., Heisler, C. A., Dopita, M. A., \& Lumsden, S. 2001, \apjs, 132, 
37

\bibitem[Kotilainen et al.(1996)]{kot96} 
Kotilainen, J. K., Moorwood, A. F. M., Ward, M. J., \& Forbes, D. A. 1996, 
\aap, 305, 107

\bibitem[Kotilainen et al.(2001)]{kot01} 
Kotilainen, J. K., Reunanen, J., Laine, S., \& Ryder, S. D. 2001, \aap, 366, 439

\bibitem[Landini et al.(1984)]{lan84} 
Landini, M., Natta, A., Oliva, E., Salinari, P., \& Moorwood, A. F. M. 1984,
\aap, 134, 284

\bibitem[Lipari et al.(2000)]{lip00} 
Lipari, S., D\'\i az, R., Taniguchi, Y., Terlevich, R., Dottori, H., \& 
Carranza, G. 2000, \aj, 120, 645

\bibitem[Lira et al.(2002)]{lira02} 
Lira, P., Ward, M. J., Zezas, A., Alonso-Herrero, A., \& Ueno, S. 2002, \mnras, 
330, 259

\bibitem[Lonsdale et al.(1984)Lonsdale, Persson, \& Matthews]{lons84} 
Lonsdale, C. J., Persson, S. E., \& Matthews, K. 1984, \apj, 287, 95

\bibitem[Mengel et al.(2001)]{meng01} Mengel, S., Lehnert, M. D., Thatte, N., 
Tacconi-Garman, L. E., \& Genzel, R. 2001, \apj, 550, 280

\bibitem[Mihos \& Hernquist(1994a)]{hos94} 
Mihos, J. C., \& Hernquist, L. 1994a, \apjl, 431, L9

\bibitem[Mihos \& Hernquist(1994b)]{hos94b} 
Mihos, J. C., \& Hernquist, L. 1994b, \apjl, 438, L47

\bibitem[Mihos \& Hernquist(1996)]{hos96} 
Mihos, J. C., \& Hernquist, L. 1996, \apj, 464, 641
 
\bibitem[Mihos et al.(1993)Mihos, Bothun, \& Richstone]{hos93} 
Mihos, J. C., Bothun, G. D., \& Richstone, D. O. 1993, ApJ, 418, 82

\bibitem[Mihos et al.(1997)]{hos97} 
Mihos, J. C., McGaugh S. S., \&  de Blok, W. J. G. 1997, \apj, 477, L49

\bibitem[Mihos et al.(1998)Mihos, Dubinski, \& Hernquist]{hos98} 
Mihos, J. C., Dubinski, J., \& Hernquist, L. 1998, \apj, 494, 183

\bibitem[Miller et al.(1997)]{mil97} 
Miller, B. W., Whitmore, B. C., Schweizer, F., \& Fall, S. M. 1997, \aj, 
114, 2381

\bibitem[Milosavljevi\'{c} \& Merritt(2001)]{milo01} 
Milosavljevi\'{c}, M., \& Merritt, D. 2001, \apj, 563, 34

\bibitem[Neff \& Ulvestad(2000)]{neff00} Neff, S. G., \& Ulvestad, J. S. 2000,
\aj, 120, 670

\bibitem[Noguchi(1988)]{nog88}
Noguchi, M. 1988, \aap, 201, 37

\bibitem[Norris \& Forbes(1995)]{nor95} 
Norris, R. P., \& Forbes, D. A. 1995, \apj, 446, 594

\bibitem[Quinlan et al.(1995)Quinlan, Hernquist, \& Sigurdsson]{quin95} 
Quinlan, G. D., Hernquist, L., \& Sigurdsson, S. 1995, \apj, 440, 554

\bibitem[Quinlan \& Hernquist(1997)]{quin97} 
Quinlan, G. D., \& Hernquist, L. 1997, New Astron., 2, 533

\bibitem[Rafanelli \& Marziani(1992)]{raf92} 
Rafanelli, P., \& Marziani, P. 1992, \aj, 103, 743

\bibitem[Read \& Ponman(1998)]{read98} 
Read, A. M., \& Ponman, T. J. 1998, \mnras, 297, 143

\bibitem[Ridgway et al.(1994)Ridgway, Wynn-Williams, \& Becklin]{ridg94} 
Ridgway, S. G., Wynn-Williams, C. G., \& Becklin, E. E. 1994, \apj, 428, 609

\bibitem[Sanders et al.(1988a)]{san88a} 
Sanders, D. B., Soifer, B. T., Elias, J. H., Madore, B. F., Matthews, K., 
Neugebauer, G., \& Scoville, N. Z. 1988a, \apj, 325, 74

\bibitem[Sanders et al.(1988b)]{san88b} 
Sanders, D. B., Scoville, N. Z., Sargent, A. I., \& Soifer, B. T. 1988b,
\apjl, 324, L55

\bibitem[Schweizer(1982)]{sch82}
Schweizer, F. 1982, \apj, 252, 455

\bibitem[Schweizer(1990)]{schw90} 
Schweizer, F. 1990, in ``Dynamics and Interactions of Galaxies'', ed. R. Wielen
(Heidelberg: Springer), 60

\bibitem[Schweizer(1996)]{schw96a} 
Schweizer, F. 1996, \aj, 111, 109

\bibitem[Schweizer(1998)]{sch98} 
Schweizer, F. 1998, in ``Galaxies: Interactions and Induced Star Formation'', 
Saas-Fee Advanced Course 26, ed. D. Friedli, L. Martinet, \& D. Pfenniger, 
(Berlin: Springer--Verlag), 105 

\bibitem[Schweizer et al.(1996)]{schw96} 
Schweizer, F., Miller, B. W., Whitmore, B. C., \& Fall, S. M. 1996, \aj, 112, 
1839

\bibitem[Scoville et al.(2000)]{sco00} 
Scoville, N. Z., et al. 2000, \aj, 119, 991

\bibitem[Springel \& White(1999)]{spr99}
Springel, V., \& White, S. D. M. 1999, \mnras, 307, 162

\bibitem[Stanford \& Balcells(1990)]{stan90} 
Stanford, S. A., \& Balcells, M. 1990, \apj, 355, 59

\bibitem[Stanford \& Balcells(1991)]{stan91b} 
Stanford, S. A., \& Balcells, M. 1991, \apj, 370, 118

\bibitem[Stanford \& Bushouse(1991)]{stan91} 
Stanford, S. A., \& Bushouse, H. A. 1991, \apj, 371, 92

\bibitem[Stauffer(1982a)]{sta82a} 
Stauffer, J. R. 1982a, \apj, 262, 66

\bibitem[Stauffer(1982b)]{sta82b} 
Stauffer, J. R. 1982b, \apjs, 50, 517

\bibitem[Surace et al.(2001)Surace, Sanders \& Evans]{sur01}
Surace, J.~A., Sanders, D.~B., \& Evans, A.~S.\ 2001, \aj, 122, 2791

\bibitem[Toomre(1977)]{tmr77} 
Toomre, A. 1977, in ``The Evolution of Galaxies and Stellar Populations,'' 
ed. B. M. Tinsley \& R. B. Larson (New Haven: Yale Univ.), 401

\bibitem[Toomre \& Toomre(1972)]{tmr72} 
Toomre, A., \& Toomre, J. 1972, \apj, 178, 623

\bibitem[Vacci et al.(1998)Vacci, Alonso-Herrero, \& Rieke]{vacci98} 
Vacci, L., Alonso-Herrero, A., \& Rieke, G. H. 1998, \apj, 504, 93

\bibitem[van der Marel \& Zurek(2000)]{marel00} van der Marel, R. P., \&
Zurek, D. 2000, in ASP Conf. Ser. 197, ``Dynamics of Galaxies: from the Early Universe to the Present'',
Ed. F. Combes, G. A. Mamon, \& V. Charmandaris, 323

\bibitem[Veilleux et al.(1995)]{vlx95} 
Veilleux, S., Kim, D.-C., Sanders, D. B., Mazzarella, J. M., \& Soifer, B. T. 
1995, \apjs, 98, 171

\bibitem[Verdoes Kleijn et al.(1999)]{ver99} 
Verdoes Kleijn, G. A., Baum, S. A., de Zeeuw, P. T., \& O'Dea, C. P. 1999, \aj, 
118, 2592

\bibitem[V\'{e}ron--Cetty \& V\'{e}ron(1986)]{ver86} 
V\'{e}ron--Cetty, M.--P., \& V\'{e}ron, P. 1986, \aap, 66, 335

\bibitem[Wang et al.(1992)Wang, Schweizer \& Scoville]{wang92}
Wang, Z., Schweizer, F., \& Scoville, N. Z. 1992, \apj, 396, 510

\bibitem[Weedman(1983)]{weed83} 
Weedman, D. 1983, \apj, 243, 756

\bibitem[Whitmore \& Schweizer(1995)]{whit95} 
Whitmore, B. C., \& Schweizer, F. 1995, \aj, 109, 960

\bibitem[Whitmore et al.(1997)]{whit97} 
Whitmore, B. C., Miller, B. W., Schweizer, F., \& Fall, S. M. 1997, \aj, 
114, 1797

\bibitem[Whitmore et al.(1999)]{whit99} 
Whitmore, B. C., Zhang, Q., Leitherer, C., Fall, S. M., Schweizer, F., 
\& Miller, B. W. 1999, \aj, 118, 1551

\bibitem[Young(1980)]{yng80} 
Young, P. J. 1980, \apj, 242, 1232

\bibitem[Yun \& Hibbard(2001)]{yun01} 
Yun, M. S., \& Hibbard, J. E. 2001, \apj, 550, 104

\bibitem[Zepf et al.(1999)]{zepf99} 
Zepf, S. E., Ashman, K. M., English, J., Freeman, K. C., \& Sharples, R. M. 
1999, \aj, 118, 752

\bibitem[Zezas et al.(2002a)]{zez02a}
Zezas, A., Fabbiano, G., Rots, A. H., \& Murray, S. S. 2002a, \apj, 577, 710

\bibitem[Zezas et al.(2002b)]{zez02b}
Zezas, A., Fabbiano, G., Rots, A. H., \& Murray, S. S. 2002b, \apjs, 142, 239

\bibitem[Zhang et al.(2001)Zhang, Fall, \& Whitmore]{zhang01} 
Zhang, Q., Fall, S. M., \& Whitmore, B. C. 2001, \apj, 561, 727

\bibitem[Zwicky(1950)]{zwc50} 
Zwicky, F. 1950, Experentia, 6, 441

\bibitem[Zwicky(1956)]{zwc56} 
Zwicky, F. 1956, Ergebnisse der Exacten Naturwissenschaften, 29, 344

\bibitem[Zwicky(1964)]{zwc64} 
Zwicky, F. 1964, \apj, 140, 1467

\end{thebibliography}
\end{document}